\documentclass{PoS}
\usepackage{textcomp,amsmath,amssymb}

\title{New theories for the Fermi scale}

\ShortTitle{New theories for the Fermi scale}

\author{\speaker{Christophe GROJEAN}%
       \\
       CERN, Physics Department, Theory Unit, Geneva, Switzerland\\
       and Institut de Physique Th\'eorique, CEA Saclay, France\\
       E-mail: \email{christophe.grojean@cern.ch}}

\abstract{Electroweak interactions need three Nambu--Goldstone bosons to provide a mass to the $W^\pm$ and the $Z$ gauge bosons but they also need an ultra-violet (UV) moderator or new physics to unitarize the gauge boson scattering amplitudes. In this talk, I will present various recent models of physics at the Fermi scale: several deformations of the Minimal Supersymmetric Standard Model, Little Higgs models, holographic composite Higgs models, 5D Higgsless models.}

\FullConference{European Physical Society Europhysics Conference on High Energy Physics\\
		 July 16-22, 2009\\
		 Krakow, Poland}

\begin{document}

\section{The Standard Model and the mass problem}
The strong, weak and electromagnetic interactions of elementary particles are described by gauge interactions based on a symmetry group $SU(3)_c\times SU(2)_L \times U(1)_Y$. Gauge theory is not only a way to classify particles and assign quantum numbers to them but it is also a dynamical principle that predicts particular couplings among particles. And the structure of these interactions has been well tested at LEP, for instance in the process $e^+e^-\to W^+W^-$. While this is certainly true at least for the 3-point functions, namely the interactions involving at least three particles, the gauge structure is actually badly violated at the level of the 2-point functions, namely in the mass spectrum: the observed mass terms for the leptons and the gauge bosons are not gauge invariant since the gauge group is chiral and also acts non-linearly on the gauge fields. This apparent clash calls for a spontaneous breaking of the gauge symmetry. 

In the broken phase, a (massive) spin one particle describes three different polarizations:  two transverse ones plus an extra longitudinal one which decouples in the massless limit. In the Standard Model (SM), the longitudinal degrees of freedom associated to the $W^\pm$ and $Z^0$ gauge bosons correspond presumably to the eaten Nambu--Goldstone bosons~\cite{Nambu:1960xd,Goldstone:1961eq} resulting from the breaking of the global chiral symmetry $SU(2)_L\times SU(2)_R/SU(2)_V$. This picture still leaves us with the question of the source of the Nambu-Goldstone bosons: What is the sector responsible for the breaking $SU(2)_L\times SU(2)_R \to SU(2)_V$? What is the dynamics of this sector? What are its interactions with the SM particles? The common lore~\cite{Englert:1964et, Higgs:1964pj} is that these extra degrees of freedom are part of a fundamental scalar field transforming as a weak doublet. This Higgs doublet corresponds to 4 real scalar fields: the 3 eaten Nambu--Goldstone  bosons and one physical real scalar degree of freedom, the notorious Higgs boson.
While this picture is in very good agreement with Electroweak (EW) data\footnote{The fact that a Higgs doublet comes with an extra approximate global symmetry (the so called custodial symmetry), which automatically ensures that the $\rho$ parameter is equal to 1 at tree level, is certainly a welcome feature in this regards.}~\cite{Barate:2003sz, Amsler:2008zzb, Hoecker:2009gd} (for a review on the Higgs boson phenomenology, see Ref.~\cite{Djouadi:2005gi}), the very fact that its unique prediction, namely the existence of the Higgs boson, has not been verified experimentally yet leaves open the possibility for other origins of the Nambu--Goldstone bosons: e.g., condensates of techniquarks, components of some gauge fields along an extra dimension \ldots

\section{The Higgs boson: a simple picture that calls for new physics}

The Higgs mechanism is at best a description, but certainly not an explanation, of electroweak symmetry breaking (EWSB) since there is no dynamics that would explain the instability of the Higgs potential at the origin. Moreover, it also jeopardizes our current understanding of the SM at the quantum level since the Higgs potential suffers from two sources of radiative instabilities: (i)~the mass term is quadratically divergent while (ii)~the quartic Higgs-self interactions could easily be driven to a Landau pole or to  a rolling vacuum at very large value of the Higgs field if the Higgs mass does not lie in the window around 130 and 170~GeV. New physics is required to solve these ``naturalness"~\cite{Weisskopf:1939, 'tHooft:1979bh, Veltman:1980mj}, ``triviality"~\cite{Wilson:1973jj,Luscher:1987ek} and ``(meta)stability"~\cite{Linde:1975sw, Weinberg:1976pe} problems. In particular the naturalness problem, also known as gauge hierarchy problem, has been the main source of inspiration/excuse for theoretical speculations on the structure of new physics above the weak scale. In the presence of generic new physics, the Higgs mass becomes UV sensitive unless a symmetry prevents it, in which radiative corrections will generate a contribution that is screened by this symmetry breaking scale and not by the UV scale of new physics. Such a symmetry should act non-linearly on the Higgs field and examples include (i)~supersymmetry (SUSY) (see Section~\ref{sec:susy}), (ii)~global symmetry when the Higgs appears as a pseudo Nambu--Goldtsone bosons (see Sections~\ref{sec:littleHiggs} and~\ref{sec:CompositeHiggs}), (iii)~gauge symmetry when the Higgs appears as a component of the gauge field along an extra spatial dimension~\cite{Manton:1979kb, Hosotani:1983xw} (see Refs.~\cite{Antoniadis:2001cv, Csaki:2002ur} for modern realizations of this idea in the context of branes and orbifold symmetry breaking and see
Ref.~\cite{Serone:2009kf} for a recent review), \ldots

There are of course additional arguments for the existence of new physics beyond the SM: (i)~at the level of $2\div3$ standard deviations, there are a few discrepancies between EW data and the SM predictions for quantities like $g_\mu-2$ or the left-right asymmetries in the hadronic and leptonic sectors; (ii)~the neutrino masses can be generated only if new states are added to the SM or if a new scale is introduced; (iii)~the SM does not provide any dynamics to generate the observed matter-antimatter asymmetry; (iv)~no SM particle can account for the dark matter (DM) relic abundance; (v)~with may be the exception of the Higgs boson itself~\cite{Bezrukov:2007ep}, no SM particle can drive inflation; (vi)~there is no rationale for the pattern of fermion masses and mixing angles; (vii)~the strong CP problem remains unexplained; (viii)~the charge quantization most likely requires an embedding of the SM gauge group into a bigger symmetry which would unify all the fundamental interactions; (ix)~gravity is left aside.

Theorists have always been very good at giving names to things they do not understand. And clearly the EWSB sector has been an inspirational source of creativity to them, as it is evident by collecting the attributes that have been associated to the Higgs boson over the last few years (for a review, see Ref.~\cite{Grojean:2007zz}): 
burried~\cite{Bellazzini:2009xt}, charming~\cite{Bellazzini:2009kw}, composite~\cite{Georgi:1981gk}, fat~\cite{Harnik:2003rs}, fermiophobic~\cite{Diaz:1994pk}, gauge~\cite{Hall:2001zb}, gaugephobic~\cite{Cacciapaglia:2006mz}, holographic~\cite{Contino:2007zz}, intermediate~\cite{Katz:2005au}, invisible~\cite{}, leptophilic~\cite{Su:2009fz}, little~\cite{ArkaniHamed:2002qx}, littlest~\cite{ArkaniHamed:2002qy}, lone~\cite{Hsieh:2008jg}, phantom~\cite{Delgado:2008px}, portal~\cite{Patt:2006fw}, private~\cite{Porto:2007ed},  slim~\cite{Wudka:1988az}, simplest~\cite{Schmaltz:2004de}, strangephilic~\cite{Lee:2009up}, twin~\cite{Chacko:2005pe}, un-~\cite{Stancato:2008mp}, unusual~\cite{Dermisek:2009si}, \ldots 
A description of these various constructions is certainly beyond the scope of this talk and I will limit myself to present a few examples that hopefully are representative of the possible structures governing the new physics needed around the Fermi scale.


\section{Supersymmetric Higgs(es)}
\label{sec:susy}

{\it A tout seigneur, tout honneur}: the SUSY extensions of the SM certainly stand out among the models of new physics for at least several reasons: (i)~the absence of quadratic divergences; (ii)~a dynamical EWSB driven radiatively by the top Yukawa interactions; (iii)~the absence of large oblique corrections due to the $R$-parity that forbids any interactions between SM fields and an odd  number of heavy fields; (iv)~a precise apparent unification of the $SU(3)\times SU(2) \times U(1)$ gauge couplings; (v)~the possibility to identify the lightest supersymmetric particle (LSP) as the Dark Matter \ldots 

Moreover, SUSY models might seem to be in even better agreement with EW data than the SM itself (this is specially true when focussing the attention to observables, like $M_W$, which prefers a light Higgs boson~\cite{Heinemeyer:2006px}). Unfortunately, this agreement is limited to small and peculiar regions of the (large) parameter space~\cite{Giudice:2006sn} and LEP2 data together with the lack of discovery of a Higgs boson below 114~GeV or of any new states has forced supersymmetry into fine-tuning territory, partially undermining its original motivation. The most stringent constraint on the scale of SUSY breaking comes from the lower bound on the Higgs boson mass. Indeed at tree-level, the strength of the Higgs (quartic) potential is dictated by gauge interactions:
\begin{equation}
V=( |\mu|^2 + m_{H_u}^2) \left| H_u^0\right|^2 + ( |\mu|^2 + m_{H_d}^2) \left| H_d^0\right|^2
-B (H_u^0 H_d^0 + h.c.) + \frac{g^2+g'^2}{8} \left(\left| H_u^0\right|^2 - \left| H_d^0\right|^2 \right)^2,   
\end{equation}
and predicts a light CP-even Higgs scalar ($\tan \beta$ measures the ratio of the two Higgs vacuum expectation values,  $v_u/v_d$):
\begin{equation}
m_h^2 = m_Z^2 \cos^2 2\beta
\end{equation}
One-loop radiative corrections can bring the Higgs mass above the LEP bound provided that a stop mass is heavier than about 1~TeV. But at the same time, the stop will also generate a correction to the $Z$ mass that should then be canceled by a large $\mu$ term. This cancelation requires a fine-tuning of the order of 1\% (the amount of fine-tuning depends exponentially on the Higgs mass bound). This ``SUSY little hierarchy problem" has driven over the years a burst of activity in building concrete models allowing for a heavier Higgs and various interesting proposals have emerged, either with a low scale of SUSY breaking mediation~\cite{Casas:2003jx} or with the addition of extra fields/symmetry/interactions to the minimal SUSY model (MSSM):
\begin{itemize}
\item more scalars: the NMSSM and its friends~\cite{Fayet:1975yi};
\item more gauge fields (with a new D-term from an asymptotically-free gauge extension)~\cite{Batra:2003nj};
\item more global symmetry: the Little SUSY models in which the SUSY Higgses appear as Goldstone bosons and benefit from a double protection~\cite{Birkedal:2004xi, Chankowski:2004mq};
\item more interactions parametrized by higher dimensional terms: the BMSSM~\cite{Brignole:2003cm, Dine:2007xi}. This effective approach offers a model-independent look at the SUSY little hierarchy problem. At the lowest order, it has been shown~\cite{Dine:2007xi} that the most general non-renormalizable interactions can be captured by the superpotential (see also Refs.~\cite{Gherghetta:1995dv, Antoniadis:2007xc, Antoniadis:2008es})
\begin{equation}
W_\textrm{\tiny BMSSM} = \frac{\lambda_1}{M} (H_u H_d)^2 + \frac{\lambda_2}{M} \mathcal{Z}_\textrm{\tiny soft} (H_u H_d)^2
\end{equation}
with no correction to the K\"ahler potential. These two interactions certainly allow for a heavier Higgs and a much lighter stop~\cite{Brignole:2003cm, Dine:2007xi, Strumia:1999jm}, while keeping the EW vacuum (meta)stable~\cite{Blum:2009na}. At the same time, the window for MSSM baryogenesis is extended and more natural~\cite{Blum:2008ym}, while the LSP can account for DM relic abundance~\cite{Bernal:2009hd}. The phenomenology of the BMSSM at colliders has been studied in Ref.~\cite{Carena:2009gx}. And implications for fine-tuning have been analyzed in Ref.~\cite{Cassel:2009ps}.
\end{itemize}

The absence of any direct evidence for SUSY prompts to keep an eye open on other possible alternatives that have emerged in the recent years with may be reduced ambition and  not necessarily aiming at providing a UV completion to the SM valid up to the Planck. All the models I will now present have an origin rooted in recent developments of string theory (branes, holography, AdS/CFT correspondence) which has provided a new laboratory to address various particle physics questions~\cite{UrangaEPS, Gherghetta:2006ha}. 

\section{Little Higgs}
\label{sec:littleHiggs}

Symmetries of the EWSB sector can help to preserve the tree-level structure of the SM, i.e., can help to keep the oblique corrections under control. For instance, it is well-known  that the embedding
\begin{equation}
\frac{SU(2)_L\times U(1)_Y}{U(1)_\textrm{\tiny em}} \subset 
\frac{SU(2)_L\times SU(2)_R}{SU(2)_V}
=\frac{SO(4)}{SO(3)}
\end{equation}
is enough to ensure that the EWSB will not generate a contribution to the $T$ oblique parameter.
The situation of the $S$ oblique parameter is notoriously more difficult to handle since the only symmetry that can protect it is the gauge $SU(2)_L$ symmetry itself or the global $SU(2)_R$ symmetry~\cite{Inami:1992rb} and they have to be broken anyway. Therefore a contribution to $S$ is expected with a scaling like $v^2/\Lambda_{EWSB}^2$, where $v=$246~GeV is the SM Higgs vacuum expectation value and  $\Lambda_{EWSB}$ is a typical mass scale of the EWSB sector. One way to reduce this contribution is to make $v$ much smaller than 
$\Lambda_{EWSB}$, which is another formulation of the gauge hierarchy problem. Again, enlarging the symmetry of the EWSB can provide a solution and ensure a naturally small contribution to the $S$ parameter: the embedding
\begin{equation}
\frac{SO(4)}{SO(3)} \subset \frac{SO(5)}{SO(4)}\ , \frac{SU(5)}{SO(5)} \ldots
\end{equation}
allows one to identify the full Higgs doublet as Nambu--Goldstone boson and leads to a naturally small ratio
$v/\Lambda_{EWSB}$.

The idea behind the Little Higgs models~\cite{ArkaniHamed:2002qy, ArkaniHamed:2001nc} is precisely to identify the Higgs doublet as a (pseudo) Nambu--Goldstone boson while keeping some sizable non-derivative interactions. By analogy with QCD where the pions $\pi^{\pm, 0}$ appear as Nambu--Godstone bosons associated to the breaking of the chiral symmetry $SU(2)_L\times SU(2)_R/SU(2)_\textrm{\tiny isospin}$, switching on some interactions that break explicitly the global symmetry will generate a mass to the would-be massless Nambu--Goldstone bosons of the order of $g \Lambda_{G/H}/(4\pi)$, where $g$ is the coupling of the symmetry breaking interaction and $\Lambda_{G/H}=4\pi f_{G/H}$ is the dynamical scale of the global symmetry 
breaking $G/H$. In the case of the Higgs boson, the top Yukawa interaction or the gauge interactions themselves will certainly break explicitly (part of) the global symmetry since they act non-linearly on the Higgs boson\footnote{When part of the global symmetry is weakly gauged, the question of alignement of the gauge group with the unbroken global symmetry arises and can give non-trivial constraints on the parameter space of the models~\cite{Grinstein:2008kt}.}. Therefore, obtaining a Higgs mass around 100~GeV would demand a dynamical scale $\Lambda_{G/H}$ of the order of 1~TeV, which is known to lead to too large oblique corrections.
Raising the strong dynamical scale by at least one order of magnitude requires an additional selection rule to ensure that a Higgs mass is generated at the 2-loop level only\begin{equation}
m_h^2 = \frac{g^2}{16 \pi^2} \Lambda_{G/H}^2 \to m_h^2= \frac{g_1^2 g_2^2}{(16 \pi^2)^2} \Lambda_{G/H}^2
\end{equation}
The way to enforce this selection rule is through a ``collective breaking" of the global symmetry:
\begin{equation}
\mathcal{L}= \mathcal{L}_{G/H} + g_1 \mathcal{L}_1 + g_2 \mathcal{L}_2.
\end{equation}
Each interaction $\mathcal{L}_1$ or $\mathcal{L}_2$ individually preserves a subset of the global symmetry such that the Higgs remains an exact Nambu--Goldstone boson whenever either $g_1$ or $g_2$ is vanishing. A mass term for the Higgs boson can be generated by diagrams involving simultaneously both interactions only. At one-loop, there is no such diagram that would be quadratically divergent. Explicitly, the cancellation of the SM quadratic divergences is achieved by a set of new particles around the Fermi scale: gauge bosons, vector-like quarks, and extra massive scalars, which are  related, by the original global symmetry, to the SM particles with the same spin. 
These new particles, with definite couplings to SM particles as dictated 
by the global symmetries of the theory, are perfect goals for the LHC.

Generically, oblique corrections in Little Higgs models are reduced either by increasing the coupling of one of the gauge group (in the case of product group models) or by increasing the mass of the $W$ and $Z$ partners, leading ultimately to a fine-tuning of the order of a few percents, i.e., improving only marginally the situation of the MSSM  (see for instance Ref.~\cite{Casas:2005ev} and references therein). The compatibility of Little Higgs models with experimental data is significantly improved when the global symmetry involves a custodial symmetry as well as a $T$-parity\footnote{There has been some doubts~\cite{Hill:2007zv} whether anomalies plague Little Higgs models with $T$ parity once gauge interactions are introduced. This is a UV-depend question and anomaly-free UV completions have been constructed~\cite{Csaki:2008se}.}~\cite{Cheng:2003ju} under which, in analogy with $R$-parity in SUSY models, the SM particles are even and their partners are odd. Such Little Higgs models would therefore appear  in colliders as jets with missing transverse energy~\cite{Carena:2006jx}. There is also interesting physics associated to the partner(s) of the top quark which could be pair-produced by gluon fusion or single-produced by $Wb$ busion, i.e., $W$-exchange in $t$-channel~\cite{Han:2003wu, Perelstein:2003wd}.

The motivation for Little Higgs models is to solve the little hierarchy problem, i.e., to push the need for  new physics (responsible for the stability of the weak scale) up to around 10~TeV. Per se, Little Higgs models are effective theories valid up to their cutoff scale $\Lambda_{G/H}$. Their UV completions could either be weakly coupled or strongly coupled (in which case, the Higgs boson appears as a true composite bound state, like the pions of QCD).

\section{Elementary vs. composite Higgs boson. Strong vs. weak EWSB}
\label{sec:strongEWSB}

What is unitarizing the $WW$ scattering amplitude? Supersymmetric models, Little Higgs models and many other models take for granted that the Higgs boson provides the answer to this pressing question of the origin of EWSB. I said earlier that the masses of the $W^\pm$ and $Z$ gauge bosons break the gauge symmetry. Actually, in the presence of these masses, the gauge symmetry is realized non-linearly: the longitudinal $W^\pm_L, Z_L$ can be described by the Nambu--Goldstone bosons, or pions,  associated to the coset 
$SU(2)_L\times SU(2)_R/SU(2)_\textrm{\tiny isospin}$ and the gauge boson mass terms correspond to the pions kinetic term ($\sigma^a$, $a=1,2,3$, are the usual Pauli matrices):
\begin{equation}
\label{eq:pions}
\mathcal{L}_\textrm{\tiny mass}= \frac{v^2}{4} \textrm{Tr} \left( D_\mu \Sigma ^\dagger D^\mu \Sigma\right)
\ \ 
\textrm{ with} \ \ 
\Sigma=e^{i \sigma^a \pi^a/v}.
\end{equation}
Thanks to this Goldstone boson equivalence~\cite{Chanowitz:1985hj}, the non-trivial scattering of the longitudinal $W$'s ($W$ generically denotes $W^\pm$ as well as $Z$) now simply follows for the contact interactions among four pions obtained by expanding the Lagrangian~(\ref{eq:pions}) and leads to amplitudes that grow with the energy:
\begin{equation}
\mathcal{A} (W_L^a W_L^b \to W_L^c W_L^d) = \mathcal{A}(s) \delta^{ab}\delta^{cd} + \mathcal{A}(t) \delta^{ac}\delta^{bd} +\mathcal{A}(u) \delta^{ad}\delta^{bc} 
\ \ 
\textrm{ with} \ \ 
\mathcal{A}(s)\approx \frac{s}{v^2}.
\end{equation}
In the absence of any new weakly coupled elementary degrees of freedom canceling this growth, perturbative unitarity will be lost around\footnote{Defining the breakdown of perturbativity is subject to arbitrary choices: the 1.2~TeV($=2\sqrt{2\pi}v$) number follows from requiring that the real part of the partial waves of the iso-amplitudes remains smaller than \textonehalf, while demanding that the tree-level amplitude remains bigger than the one-loop one leads to the more conventional scale, $4 \pi v(\approx$~3.1~TeV), associated to a non-linear $\sigma$-model with  a breaking scale $v$.} 1.2~TeV and new strong dynamics will kick in and soften the UV behavior of the amplitude, for instance via the exchange of massive bound states similar to the $\rho$ meson of QCD. In any circumstances, by measuring the $W^\pm$ and $Z$ masses, we have been guaranteed to find new physics around the Fermi scale to ensure the proper decoupling of the longitudinal polarizations  at very high energy.

The simplest  example of new dynamics that can restore perturbative unitarity consists of a single scalar field, $h$, singlet under $SU(2)_L\times SU(2)_R/SU(2)_V$ and coupled to the longitudinal $W$'s as~\cite{SILH2}:
\begin{equation}
	\label{eq:scalar}
\mathcal{L}_\textrm{\tiny EWSB}= \frac{1}{2} \partial_\mu h \partial^\mu h - V(h) + \frac{v^2}{4} \textrm{Tr} \left( D_\mu \Sigma ^\dagger D^\mu \Sigma\right) \times \left(1+ 2 a \frac{h}{v} + b \frac{h^2}{v^2} \right).
\end{equation}
Via its linear coupling, $a$, to the $W_L$'s, the scalar gives an additional contribution to the $WW$ scattering amplitude
\begin{equation}
\mathcal{A}_\textrm{\tiny scalar exchange}(s) = - \frac{a^2\, s^2}{v^2(s-m_h^2)},
\end{equation}
which, for $a=1$, cancels the leading contact term at high energy. This is not the end of the story yet: perturbative unitarity should be maintained in inelastic channels too, like $W_LW_L \to hh$. Both the linear and quadratic couplings, $a$ and $b$, contribute to this amplitude and the terms growing with the energy are canceled for the particular choice $b=a^2$. The point $a=b=1$ defines the SM Higgs boson and it can be shown that the scalar resonance and the pions then combine together to form a doublet transforming {\it linearly} under $SU(2)_L\times SU(2)_R$.

The Lagrangian~(\ref{eq:scalar}) describes either an elementary or a composite Higgs boson. As soon as the couplings deviate from $a=b=1$, the Higgs exchange alone will fail to fully unitarize the $WW$ scattering amplitude irrespectively whether or not the effective Lagrangian ~(\ref{eq:scalar}) emerges from a perturbative theory (see for instance Refs.~\cite{Bellazzini:2009xt, Bellazzini:2008zy}) or from a strongly interacting dynamics. Therefore and contrary to a general belief, the question of  strong vs weak dynamics at the origin of the EWSB is decoupled from the question of the existence of a light and narrow Higgs-like scalar. In composite Higgs models, the deviations from $a=b=1$ are controlled (see Section~\ref{sec:CompositeHiggs}) by the ratio of the weak scale over the Higgs compositeness scale, $f$, which can be rather low (a few hundreds of~GeV), and strong $WW$ scattering above the Higgs mass is therefore expected.

Composite Higgs models are examples of models where the breakdown of perturbative unitarity is postponed to higher energy\footnote{A famous theorem due to Cornwall et al.~\cite{Cornwall:1973tb} states that the only way to fully ensure perturbative unitarity in all possible elastic and inelastic channels is via the exchange of a Higgs boson in a spontaneously broken gauge theory. We explore here the possibility to delay the perturbative unitarity breakdown.}. In the next section, I will discuss in details the collider signatures of these models. The extra dimensional Higgsless models that will be presented in Section~\ref{sec:Higgsless} are other examples with delayed perturbative unitarity breakdown thanks to a non-trivial dynamics of a tower of spin 1 massive particles. Both classes of models were already considered in the eighties (see for instance Ref.~\cite{Hill:2002ap} for a historical perspective,  and Ref.~\cite{Sannino:2008kg} for an account on recent developments) but they have experienced a recent revival and can now be cast in terms of models with warped extra dimensions.  Thanks to this holographic description,  models of strong EWSB can now be extrapolated to the far UV up to energies of the order of the Planck scale and questions like the unification of gauge couplings can be legitimately addressed (see, for instance, Ref.~\cite{Agashe:2005vg}). Overall, serious competitors to the MSSM have emerged as possible extension of the SM at the Fermi scale.

\begin{figure}[htbp]
\begin{center}
\includegraphics[width=.75\textwidth]{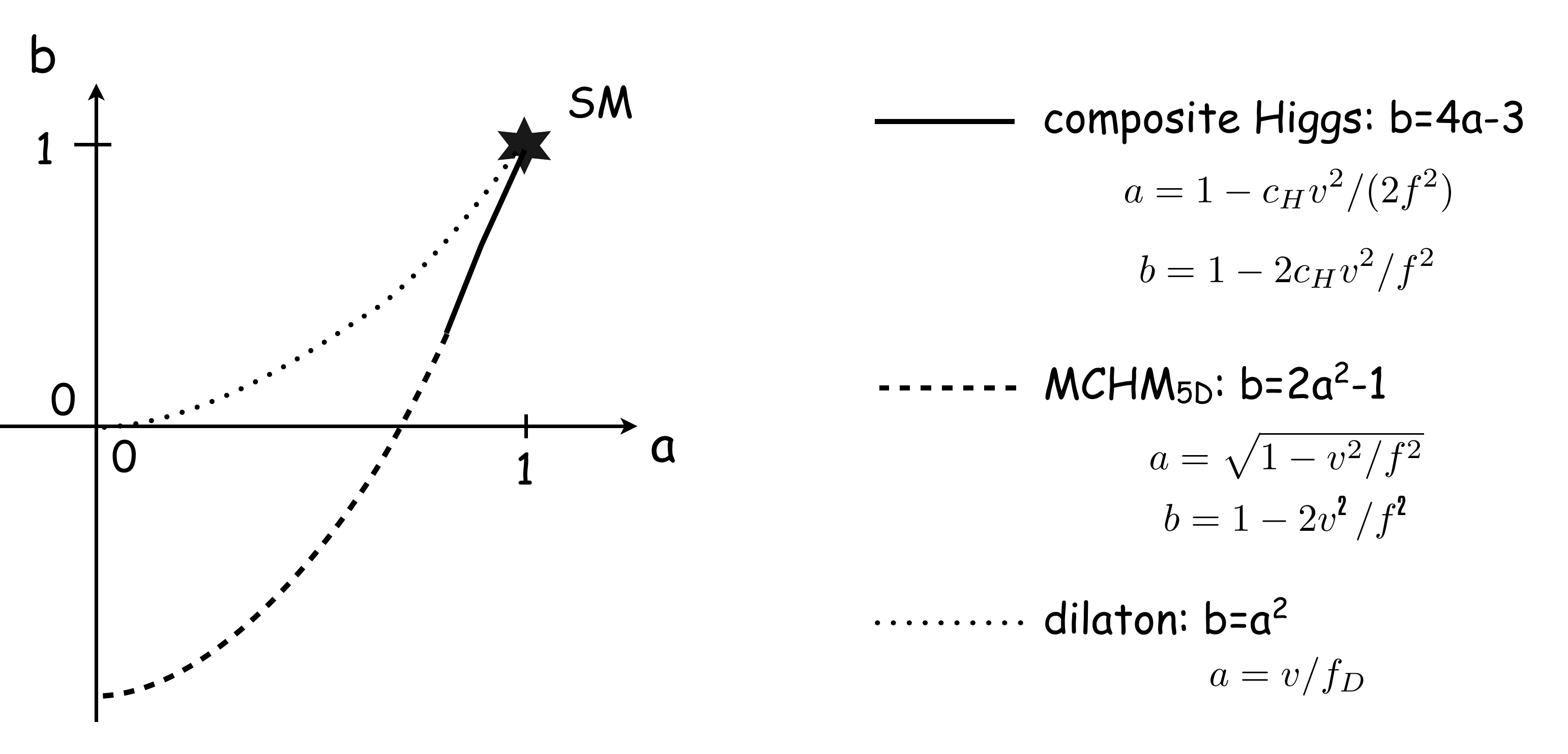}
\caption[]{(Partial) unitarization of scattering amplitudes with a scalar field. 
$a$ and $b$ are the linear and quadratic couplings of the scalar to the gauge bosons defined in
Eq.~(\ref{eq:scalar}). The particular point $a=b=1$ corresponds to the SM and neither the elastic nor the inelastic scattering amplitudes are growing with the energy. Composite Higgs models with a rather large compositeness scale, $f\gg v$, are sitting on a universal line away from the SM (this universal behavior is the result of the fact that the scattering amplitudes are dominated by a single operator, as I will show in Section~\ref{sec:CompositeHiggs}). When the compositeness scale is lowered towards the weak scale, the series in $v/f$ has to be resumed and model dependent effects appear. For instance, the minimal composite Higgs model (MCHM$_{5D}$) of Ref.~\cite{Agashe:2004rs} sits along the curve $b=2a^2-1$. It was shown in Ref.~\cite{Low:2009di} that, under very generic assumptions, in composite Higgs models, $a$ is always smaller than 1. The dilaton couplings can also be described by the Lagrangian~(\ref{eq:scalar}) and they satisfy $b=a^2$, therefore the double dilaton production never becomes strong.
}
\label{fig:scalar}
\end{center}
\end{figure}

\section{(Pseudo Nambu--Goldstone) composite Higgs models}
\label{sec:CompositeHiggs}

Notwithstanding its simplicity, the appeal of the SM Higgs picture comes from its successful agreement with EW precision data, provided that the Higgs boson is rather light. In this regard, being an elementary scalar is not a virtue but rather a flaw because of the quadratic divergence destabilizing the Higgs mass. It is thus tantalizing to consider the Higgs boson as a composite bound state emerging from a strongly-interacting sector. In order to maintain a good agreement with EW data, it is sufficient that a mass gap separates the Higgs resonance from the other resonances of strong sector (the resonances that will ultimately enforce a good behavior of the $WW$ scattering amplitudes). Such a mass gap can naturally follow from dynamics if the strongly-interacting sector possesses a global symmetry, $G$, spontaneously broken at a scale $f$ to a subgroup $H$, such that the coset $G/H$ contains a fourth Nambu--Goldstone bosons that can be identified with the Higgs boson.  Simple examples of such coset are $SU(3)/SU(2)$ or $SO(5)/SO(4)$, the latter being favored since it is invariant under the custodial symmetry (some non-minimal models with extra Nambu--Goldstone bosons have also been constructed~\cite{Gripaios:2009pe}). Attempts to construct composite Higgs models in 4D have been made by Georgi and Kaplan (see for instance Ref.~\cite{Georgi:1984af}) and modern incarnations  have been recently investigated in the framework of 5D warped models where, according to the principles of the AdS/CFT correspondence, the holographic composite Higgs boson now originates from a component of a gauge field along the 5th dimension with appropriate boundary conditions\footnote{A Higgs localized on the IR brane of a warped model, like in the original models of Randall--Sundrum~\cite{Randall:1999ee}, would also correspond to a 4D composite Higgs boson in the dual interpretation~\cite{ArkaniHamed:2000ds, Rattazzi:2000hs}. However, this composite Higgs state would not be a pseudo Nambu--Goldstone boson and there is no reason why it would be lighter than the other resonances of the strong sector (in other words, the original Randall--Sundrum models solve the gauge hierarchy problem but they do not address the little hierarchy problem. Hence the need to rely on the $A_5$ degree of freedom.}.  

The composite Higgs models offer a nice  and continuous interpolation between the SM and technicolor type models. The dynamical scale $f$ defines the compositeness scale of the Higgs boson: when $\xi=v^2/f^2 \to 0$, the Higgs boson appears essentially as a light elementary particle (and its couplings approach the ones predicted by the SM) while the other resonances of the strong sector become heavier and heavier and decouple; on the other hand, when $\xi \to 1$, the couplings of the Higgs boson to the $W_L$'s go to zero and unitarity in gauge boson scattering is ensured by the exchange of the heavy resoances.

At the eve of the LHC operation, I would like to give a description of the physics of such a composite Higgs boson rather than presenting the details of the construction of an explicit model. In the same way that we do not need  the refinements of QCD to describe the physics of the pions, I will rely on an effective Lagrangian to capture the relevant physics. This effective Lagrangian involves higher dimensional operators for the low energy degrees of freedom (the SM particles and a unique Higgs boson in the minimal case) and the strong sector will be broadly parametrized by two quantities: the typical mass scale, $m_\rho$, of the heavy resonances and the dynamical scale, $f$, associated to the coset $G/H$ (for maximally strongly coupled sectors, we expect $m_\rho \approx 4 \pi f$; here, I will simply assume that $m_\rho$ is parametrically larger than $f$). There are two classes of higher dimensional operators: (i)~those that are genuinely sensitive to the new strong force  and will affect qualitatively the physics of the Higgs boson and (ii)~those that are sensitive to the spectrum of the resonances only and  will simply act as form factors. Simple rules control the size of these different operators, see Ref.~\cite{Giudice:2007fh},  and the effective Lagrangian generically takes the form ($g, g'$ are the SM EW gauge couplings, $\lambda$ is the SM Higgs quartic coupling and $y_f$ is the SM Yukawa coupling to the fermions $f_{L,R}$)\footnote{This effective Lagrangian captures the physics of a Higgs boson identified as pseudo Nambu--Goldstone boson emerging from a {\it strongly} interacting sector. The scaling of the operators will be different in perturbative theories. For instance, in Little Higgs models with a product group gauge symmetry, there will be sizable corrections to the non-linear $\sigma$-model structure when none of the gauge couplings is strong. The Higgs potential, in the case of a strongly interacting light Higgs (SILH), is also fully saturated by quantum effects, i.e., generated radiatively, while in Little Higgs models,  the quartic interactions can be larger~\cite{Giudice:2007fh}.}:
\begin{eqnarray}
&&\mathcal{L}_{\rm SILH} = \frac{c_H}{2f^2} \left( \partial_\mu \left( H^\dagger H \right) \right)^2
+ \frac{c_T}{2f^2}  \left(   H^\dagger{\overleftrightarrow D}_\mu H\right)^2 
- \frac{c_6\lambda}{f^2}\left( H^\dagger H \right)^3 
+ \left( \frac{c_yy_f}{f^2}H^\dagger H  {\bar f}_L Hf_R +{\rm h.c.}\right) \nonumber \\ 
&&
+\frac{ic_Wg}{2m_\rho^2}\left( H^\dagger  \sigma^i \overleftrightarrow {D^\mu} H \right )( D^\nu  W_{\mu \nu})^i
+\frac{ic_Bg'}{2m_\rho^2}\left( H^\dagger  \overleftrightarrow {D^\mu} H \right )( \partial^\nu  B_{\mu \nu})  +\ldots 
\label{eq:silh}
\end{eqnarray}
All the coefficients, $c_H, c_T \ldots$, appearing in Eq.~(\ref{eq:silh}) are expected to be of order one.

\begin{figure}[!th]
\begin{center}
\includegraphics[width=0.45\textwidth,clip,angle=0]{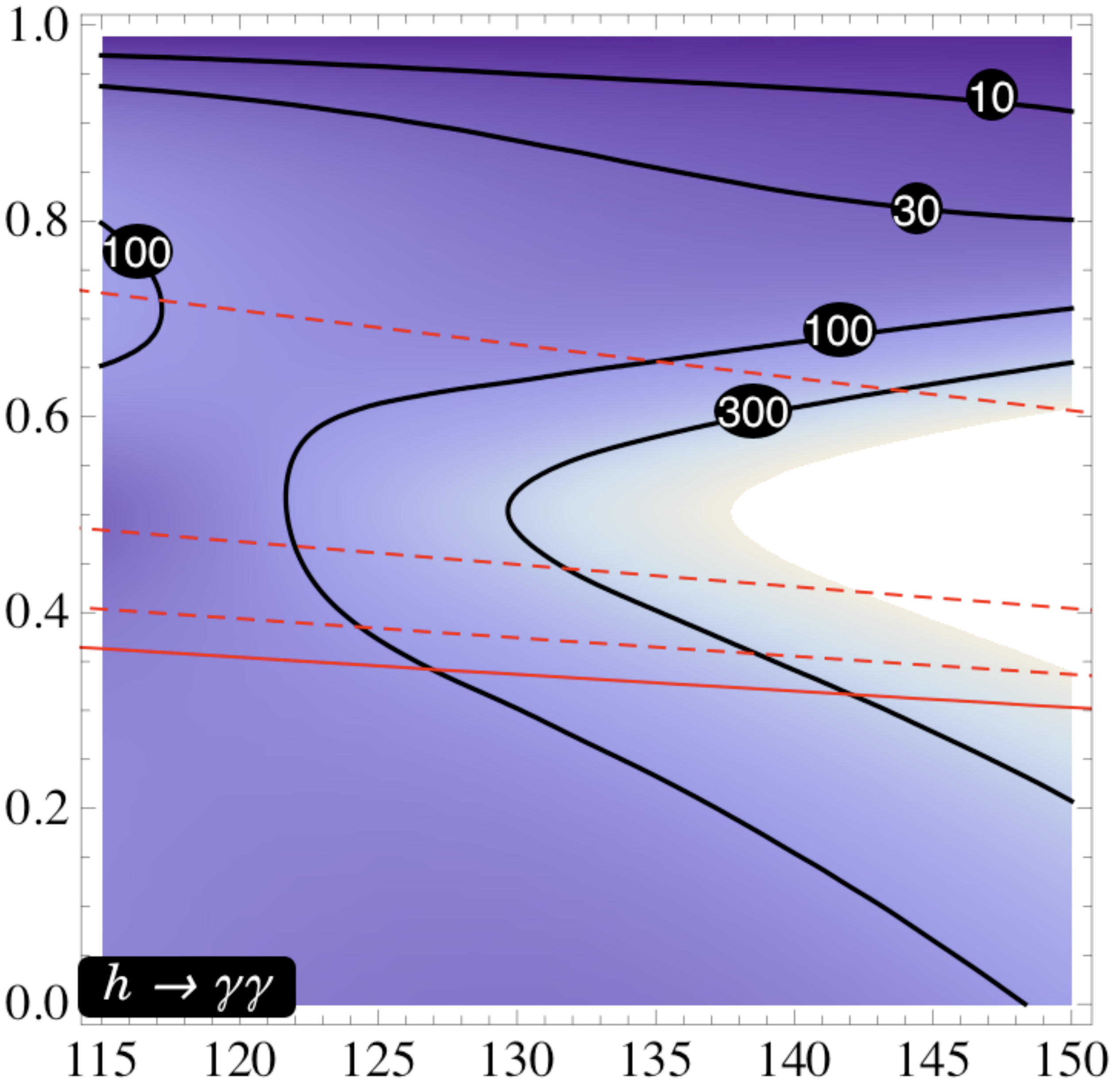}
\hspace*{0.5cm}
\includegraphics[width=0.45\textwidth,clip,angle=0]{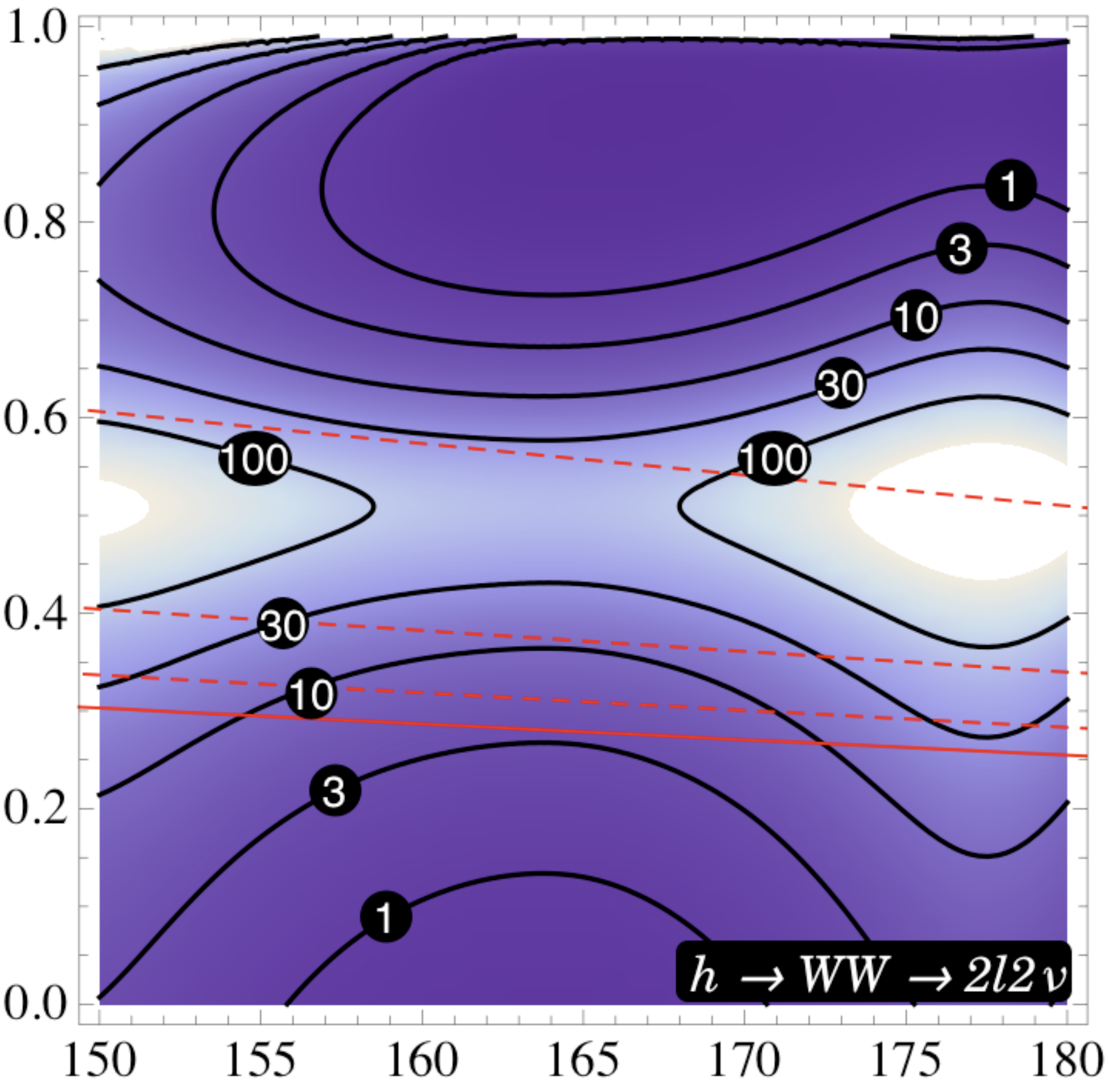}\\[.2cm]
\includegraphics[width=0.45\textwidth,clip,angle=0]{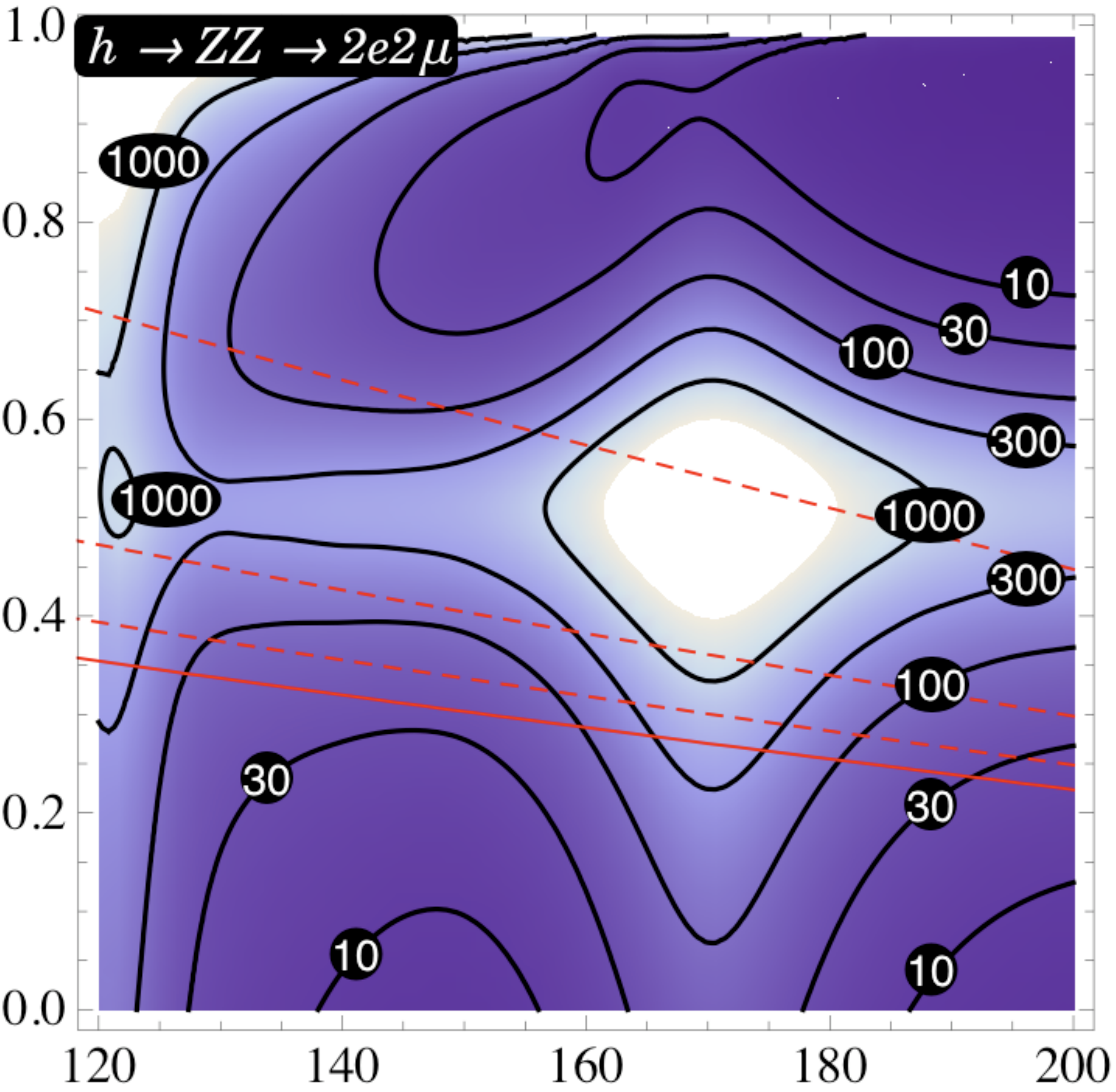}
\hspace*{0.5cm}
\includegraphics[width=0.45\textwidth,clip,angle=0]{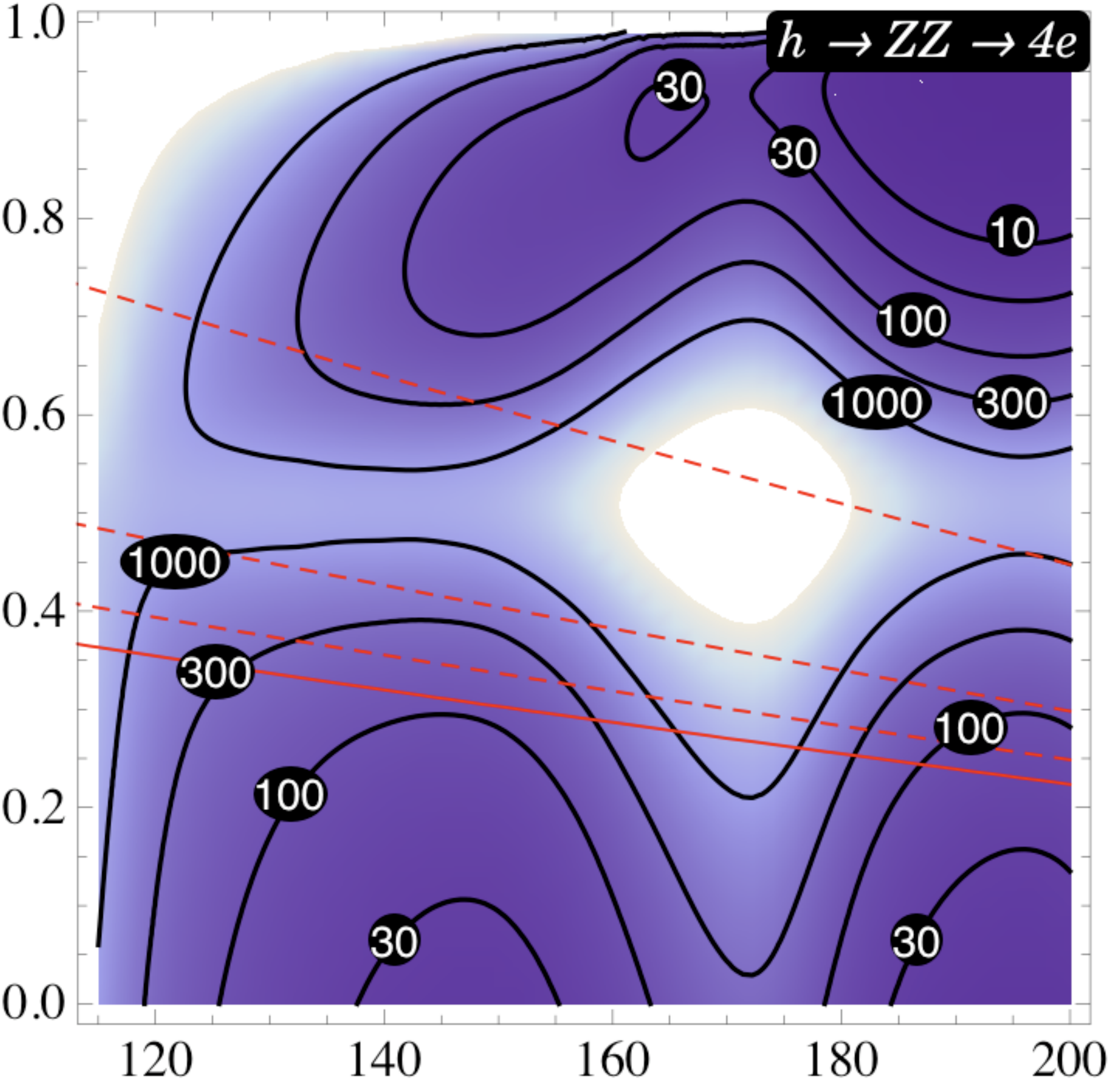}\\
\caption{ \label{fig:HiggsSearches}
Luminosity (in fb$^{-1}$) needed for discovery in the most promising channels with the CMS detector~\cite{Ball:2007zza} as a function of the Higgs mass (horizontal axis, in~GeV) and the parameter $\xi=v^2/f^2$ (vertical axis), measuring the amount of compositeness of the Higgs boson. These contour plots correspond to the minimal composite Higgs model of Ref.~\cite{Agashe:2004rs} with fermion in fundamental representations. 
Strictly speaking, these results are fuly consistent only below the plain red curve that delineates the 99\% C.L. region of the parameters space favored by EW precision constraints (with a cutoff scale fixed at 2.5~TeV). Above, some amount of cancellation in the oblique parameters is needed (from the bottom to the top, the dashed red lines indicate regions with 10\%, 25\% and 50\% cancellation in $S$ and $T$), and the physics at the origin of this cancellation might also affect the Higgs production cross sections and decay rates.
From Ref.~\cite{SILH3}. 
}
\end{center}
\end{figure}

Some oblique corrections are generated, at tree-level, by the operators of this effective Lagrangian: (i)~the operator $c_T$ gives a contribution to the $T$ Peskin--Takeuchi parameter, $\hat{T}=c_Tv^2/f^2$, which would impose a very large compositeness scale; however, assuming that the   custodial symmetry is preserved by the strong sector,   the coefficient of this operator is vanishing automatically; (ii)~a contribution to the $S$ parameter is generated by the form factor operators only, $\hat{S}=(c_W+c_B) m_W^2/m_\rho^2$, and will simply impose a lower bound on the mass of the heavy resonances, $m_\rho \geq 2.5$~TeV. At the loop level, the situation is getting a bit more complicated: as I am going to show below, the couplings of the Higgs to the SM vectors receive some corrections of the order $v^2/f^2$, and these corrections prevent the nice cancelation occurring in the SM between the Higgs and the gauge boson contributions and $S$ and $T$ are logarithmically divergent~\cite{Barbieri:2007bh}  (the divergence in $T$ will enventually be screened by resonance states if the strong sector is invariant under the custodial symmetry). Typically, this one-loop IR contribution imposes~\cite{Anastasiou:2009rv, Contino:2009ez} $f^2/v^2 \geq 3 \div 4$, as reported on Fig.~\ref{fig:HiggsSearches} (see also Refs.~\cite{Agashe:2005dk, Carena:2007ua, Gillioz:2008hs, Bouchart:2008vp} for careful discussions of electroweak precision tests in composite models built in 5D). Overall, $\xi=v^2/f^2$ is a good estimate of the amount of fine-tuning of these models~\cite{Contino:2006qr}.

One may worry that because of the modified Yukawa interactions induced by the operator $c_y$, the mass matrices and the Yukawa interaction matrices  are not simultaneously diagonalizable, leading to potentially dangerous flavor changing neutral currents (FCNC). Actually, the coefficient $c_y$ is flavor universal (at least among the light fermions) and the flavor structure of the higher dimensional Yukawa interactions are proportional to the  SM Yukawa interactions. In other words, the effective Lagrangian~(\ref{eq:silh}) satisfies the minimal flavor violation hypothesis~\cite{D'Ambrosio:2002ex}. And therefore no Higgs-mediated FCNC is generated at the leading level (see however Refs.~\cite{Agashe:2009di, Azatov:2009na} for a detailed discussion on this subject). Besides the Higgs boson, the other resonances of the strong sector can generate too large flavor violating amplitude. The extra dimensional realizations of the composite Higgs models provide some clues on this issue. In these setups, the hierarchy among the fermion masses and mixing angles is the result of flavor dependent wavefunctions~\cite{ArkaniHamed:1999dc, Grossman:1999ra} (the dual interpretation is that the light fermions are only {\it partially} composite~\cite{Kaplan:1991dc}). At the same time, a built-in RS--GIM mechanism~\cite{Huber:2003tu, Agashe:2004ay} highly suppress the FCNC processes and the KK scale, equivalent to $m_rho$, can be lowered to 20~TeV~\cite{Csaki:2008zd} or even to 5$\div$6~TeV~\cite{Agashe:2008uz} (to be compared to the generic 10$^{4\div 5}$~TeV scale in the absence of any suppression), the stringent constraint actually coming for CP violation in the kaon sector~\cite{Buras:2009if}.

The effective Lagrangian~(\ref{eq:silh}) does induce some corrections to the Higgs couplings to the SM particles. In particular, the operator $c_H$ gives a correction to the Higgs kinetic term which can be brought back to its canonical form at the price of a proper rescaling of the Higgs field inducing an universal shift of the Higgs couplings by a factor $1-c_H\, v^2/(2f^2)$. For the fermions, this universal shift adds up to the modification of the Yukawa interactions:
\begin{eqnarray}
&&g_{hf\bar{f}}^{\xi} = g_{hf\bar{f}}^\textrm{\tiny SM}\times (1-(c_y + c_H/2) v^2/f^2),\\
&&g_{hWW}^{\xi} = g_{hWW}^\textrm{\tiny SM} \times (1-c_H\, v^2/(2f^2)).
\end{eqnarray}
All the dominant corrections, i.e. the ones controlled by the strong operators, preserve the Lorentz structure of the SM interactions, while the form factor operators will also introduce couplings with a different Lorentz structure.
 
\begin{figure}[t]
\begin{center}
\includegraphics[width=0.45\textwidth,clip,angle=0]{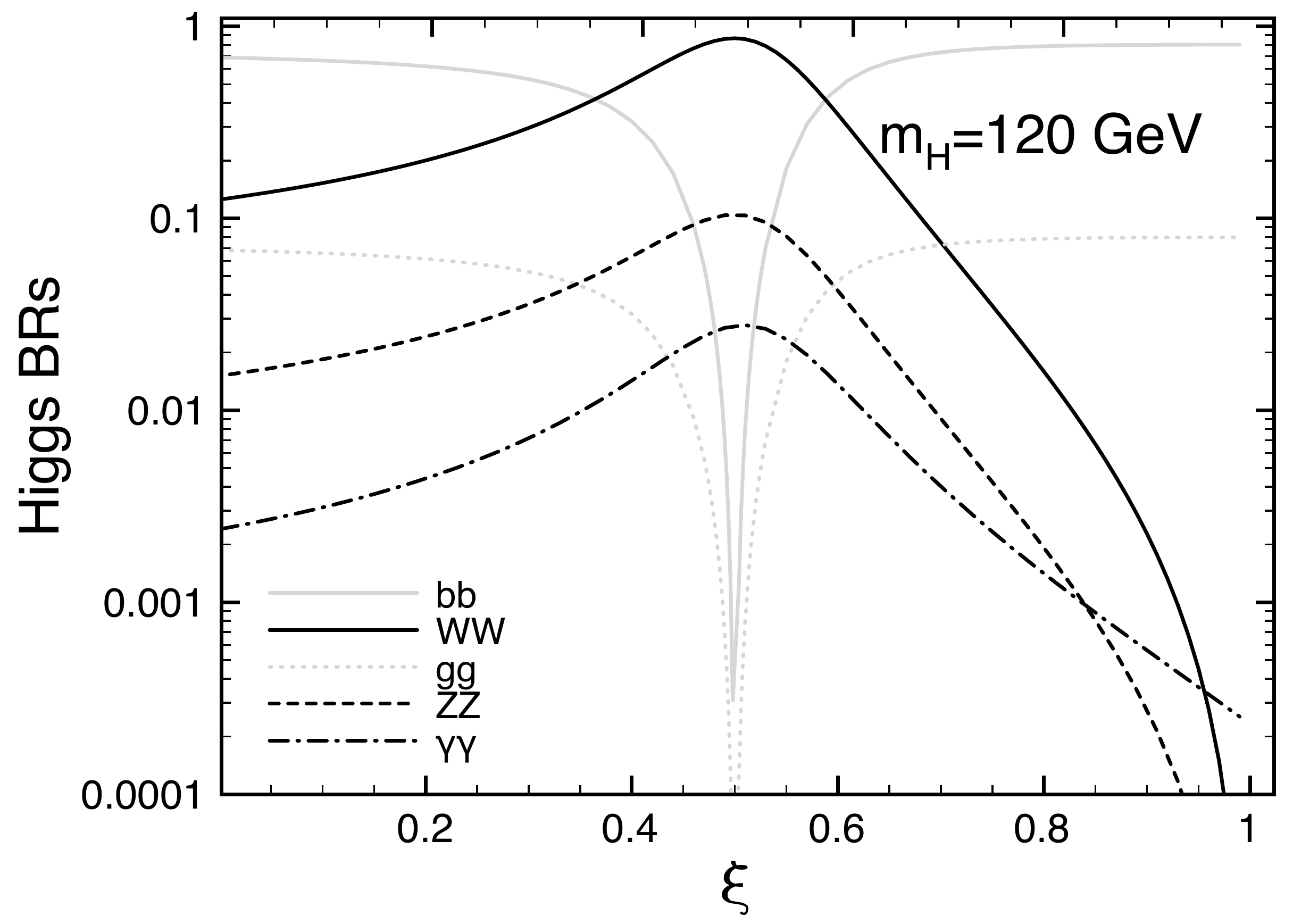}
\includegraphics[width=0.45\textwidth,clip,angle=0]{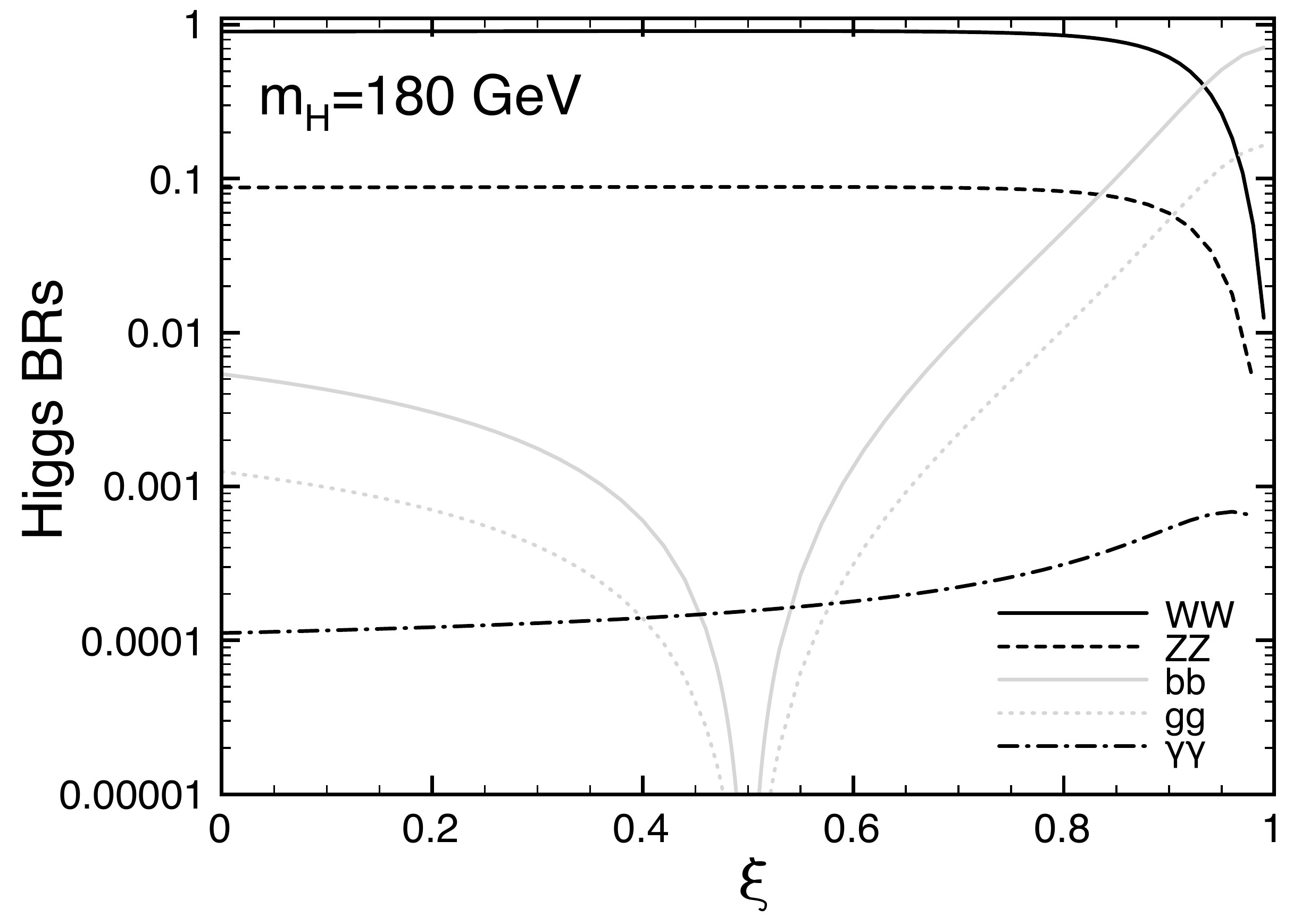}
\caption{ \label{fig:BRs}
Higgs decay branching ratios as a function of $\xi=v^2/f^2$ for SM fermions embedded into fundamental 
representations of $SO(5)$ for two benchmark Higgs masses:
$m_h=120$~GeV (left plot) and 
$m_h=180$~GeV (right plot). 
For $\xi=0.5$, the Higgs is fermiophobic, while in the technicolor limit, 
$\xi \to 1$, the Higgs becomes gaugephobic. From Ref.~\cite{SILH2}.
}
\end{center}
\end{figure}

The effective Lagrangian~(\ref{eq:silh}) should be viewed as the first terms in an expansion in $\xi=v^2/f^2$. When departing significantly from the SM limit, $v^2/f^2 \sim \mathcal{O}(1)$, the series has to be resummed. Explicit models, like the ones constructed in 5D warped space~\cite{Agashe:2004rs}, provide examples of such a resummation, allowing to study the effects of the anomalous Higgs couplings up to the technicolor limit. Figure~\ref{fig:BRs} shows the modification in the branching ratios for the Higgs decays to SM particles in the minimal composite Higgs model with fermions embedded into fundamental representations of $SO(5)$.

The Higgs anomalous couplings affect the decay rates as well as the production cross sections of the Higgs~\cite{Giudice:2007fh, Djouadi:2007fm, Falkowski:2007hz}. Therefore, the searches for the Higgs boson at the LHC, as well as the LEP/Tevatron exclusion bounds are modified as compared to the SM case. Figure~\ref{fig:HiggsSearches} reports the amount of luminosity needed for discovery  in the most promising channels for the minimal composite Higgs models of Ref.~\cite{Agashe:2004rs}.

Will the LHC be able to probe these deviations in the couplings\footnote{The physics of the composite models, as captured by the effective Lagrangian~(\ref{eq:silh}), selects the operators $c_H$ and $c_y$ as the most important ones for LHC studies, as opposed to totally model-independent operator analyses~\cite{Manohar:2006gz, Pierce:2006dh} which often lead to the conclusion that the dominant effects should appear in the vertices $h\gamma \gamma$ and $hgg$, since their SM contribution occurs only at loop level.} of the Higgs?  The contribution of the operator $c_H$ is universal for all Higgs couplings and therefore it does not affect the Higgs branching ratios, but only the total decay width and the production cross section. The measure of the Higgs decay width at the LHC is very difficult and it can be reasonably done only for rather heavy Higgs bosons, well above the two gauge boson threshold, 
a region which is not of particular interest since we consider  the Higgs as a pseudo-Goldstone boson, and therefore relatively light. However, for a light Higgs, LHC experiments can measure the product $\sigma_h \times BR_h$ in many different channels: production through gluon, gauge-boson fusion, and top-strahlung; decay into $b$, $\tau$, $\gamma$ and (virtual) weak gauge bosons.
At the LHC with about 300~fb$^{-1}$, it is possible to measure Higgs production rate times branching ratio in the various channels with 20--40~\% precision~\cite{Duhrssen:2003, Duhrssen:2004cv}. For $c_H$ and $c_y$ of order one, this will translate into a sensitivity on the compositeness scale of the Higgs, $4\pi f$, up to $5\div 7$~TeV.
  
Cleaner experimental information can be extracted from ratios between the rates of processes with the same Higgs production mechanism, but different decay modes. In measurements of these ratios of decay rates, many systematic uncertainties drop out. However, the Higgs coupling determinations at the LHC will still be limited by statistics, and therefore they can benefit from a luminosity upgrade, like the sLHC~\cite{Gianotti:2002xx}. At a linear collider, like an ILC operating at $\sqrt{s}=500$~GeV~\cite{Djouadi:2007ik}, the precision on $\sigma_h \times BR_h$ can reach the percent level~\cite{AguilarSaavedra:2001rg}, providing a very sensitive probe on the compositeness scale of the Higgs up to $4\pi f\sim 30$~TeV. Moreover, a linear collider can test the existence of the operator $c_6$ that controls the Higgs self interactions, since the triple Higgs coupling can be measured with an accuracy of about 10\% for $\sqrt{s} =500$~GeV and an integrated luminosity of 1~ab$^{-1}$~\cite{Barger:2003rs}.  CLIC running at $\sqrt{s}=3$~TeV~\cite{Accomando:2004sz}, will improve these sensitivities by a factor 2~\cite{Barger:2003rs}.

Deviations from the SM predictions of Higgs production and decay rates could be a hint towards models with strong dynamics, especially if no new light particles are discovered at the LHC. However, they do not unambiguously  imply the existence of a new strong interaction. The most characteristic signals of a composite Higgs model have to be found in the very high-energy regime. Indeed, as already announced in Section~\ref{sec:strongEWSB}, a peculiarity of a composite Higgs boson is that it fails to fully unitarize the $W_LW_L$ scattering amplitudes which have thus a residual growth with energy and the corresponding interaction becomes strong, eventually violating tree-level unitarity at the cutoff scale. Indeed, the extra contribution to the Higgs kinetic term from the  $c_H$ operator prevents Higgs exchange diagrams  from accomplishing the exact cancellation, present in the SM, of the terms growing with energy in the amplitudes. Therefore, although the Higgs is light, we obtain strong $WW$ scattering at high energies.

From the operator $c_H$, using the Goldstone equivalence theorem, it is easy to derive the following high-energy limit of the scattering amplitudes for longitudinal gauge bosons
\begin{equation}
	\label{eq:SILHamplitudes}
\mathcal{A} (W_L^a W_L^b \to W_L^c W_L^d) = \mathcal{A}(s) \delta^{ab}\delta^{cd} + \mathcal{A}(t) \delta^{ac}\delta^{bd} +\mathcal{A}(u) \delta^{ad}\delta^{bc} 
\ \ 
\textrm{ with} \ \ 
\mathcal{A}(s)\approx \frac{s}{f^2}.
\end{equation}
This result is correct to leading order in $s/f^2$, and to all orders in $v^2/f^2$ in the limit $g_\textrm{\tiny SM}=0$, when the $\sigma$-model is exact. 
The growth with energy of the amplitudes is strictly valid only up to the maximum energy of our effective theory, namely $m_\rho$. The behaviour above $m_\rho$ depends on the specific model realization. In 5D models,  the growth of the elastic amplitude is softened by Kaluza--Klein modes exchange~\cite{Falkowski:2007iv}, but the inelastic channels dominate and strong coupling is reached at a scale $\sim 4\pi f$.
Notice that the amplitudes~(\ref{eq:SILHamplitudes}) are exactly proportional to the scattering amplitudes obtained in a Higgsless SM, the growth being controlled by the strong coupling scale, $f$, and not the weak scale itself, $v$. 

\begin{figure}[t]
\begin{center}
\includegraphics[width=0.45\textwidth,clip,angle=0]{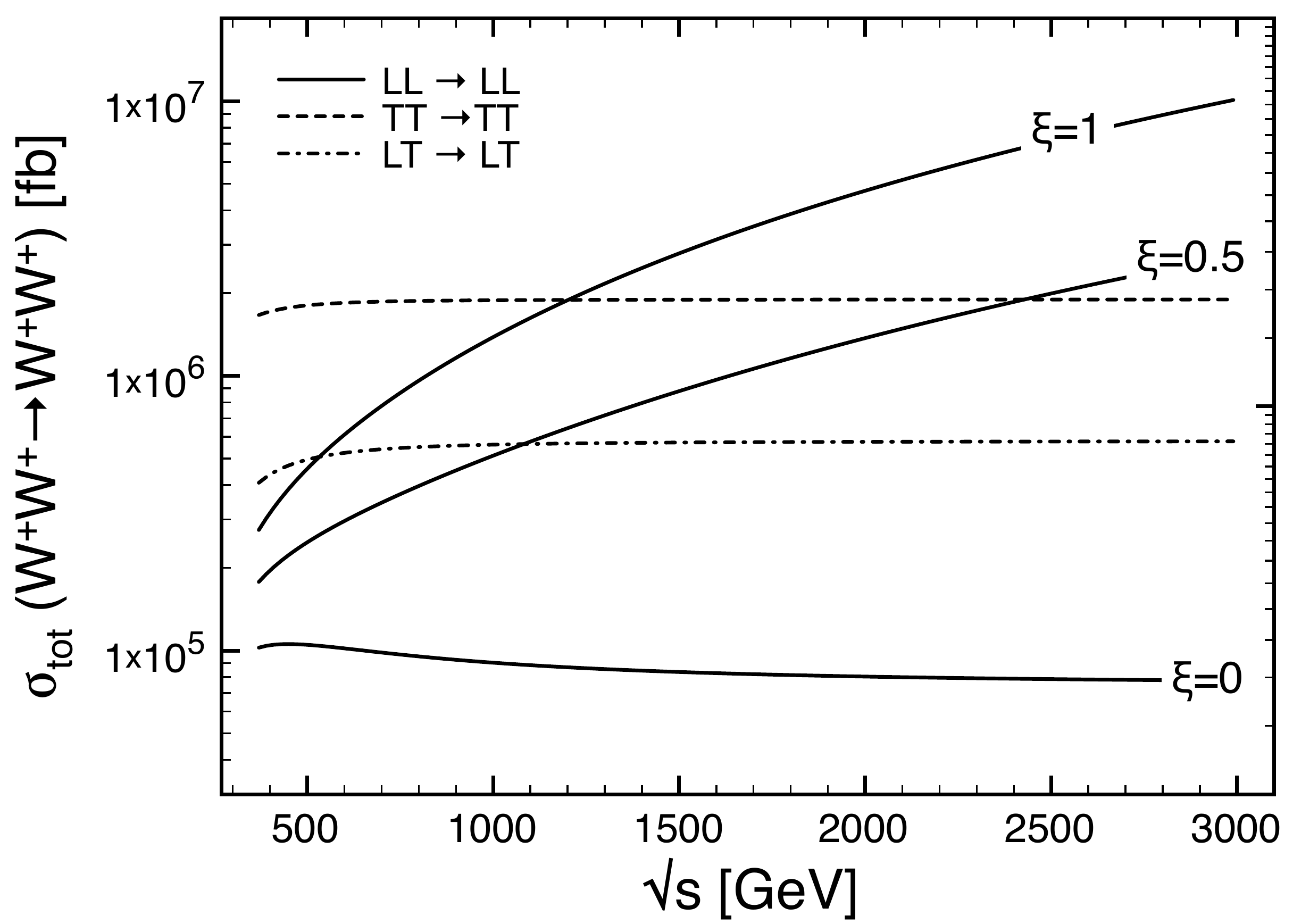}
\hspace{.2cm}
\includegraphics[width=0.45\textwidth,clip,angle=0]{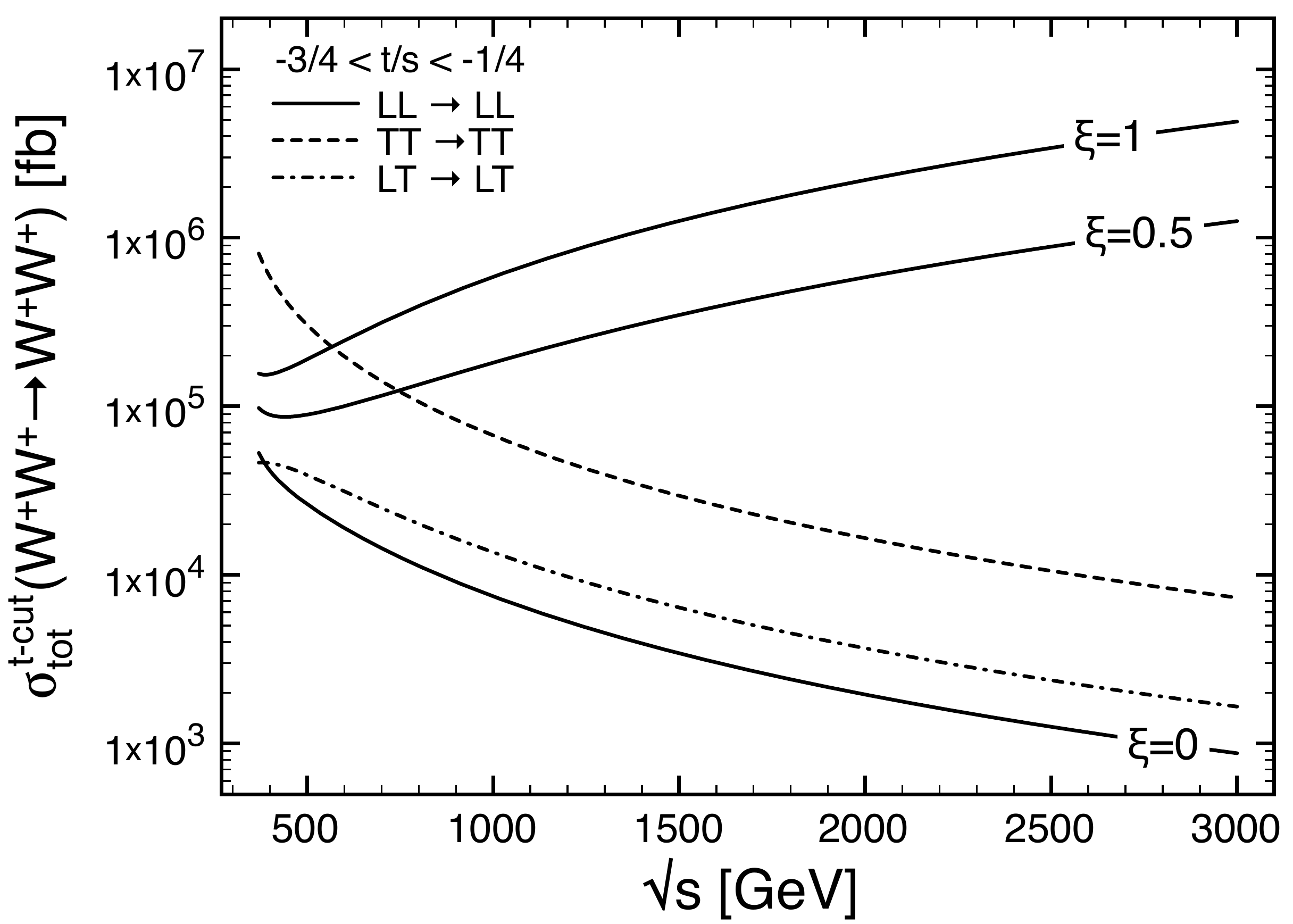}
\caption{\label{fig:WpWpTOWpWp}
 $W^+W^+\to W^+W^+$ partonic cross section as a function of the center of mass energy for $m_h = 180$~GeV for the SM ($\xi=0$) and for composite Higgs models ($\xi=v^2/f^2 \not =0$).
On the left, the inclusive cross section (with a cut on $t$ and $u$ of the order of the 
$W$ squared mass) is shown. The right plot  presents the hard cross section with a cut $-0.75 < t/s < -0.25$.  
Within the SM, the scattering cross section of the longitudinal polarizations is over-dominated by the one of the transverse polarizations. Therefore the onset of strong scattering in composite Higgs models is delayed to energies much higher than $M_W$, as estimated by NDA.
The enhancement of the amplitude for the transverse polarization is due to the structure of the Coulomb terms for the exchange of the photon and the $Z$ in the $t$- and $u$-channels. From Ref.~\cite{SILH2}.
}
\end{center}
\end{figure}

Will the LHC be able to measure the growth of these scattering amplitudes? Contrary to a naive belief, it is a notoriously difficult measurement which requires some large integrated luminosity~\cite{Bagger:1995mk}. The most promising channels correspond to purely leptonic decays of the $W$'s, though semileptonic decay channels have also been considered recently~\cite{Butterworth:2002tt, Ballestrero:2009vw}. The rapid falloff of the $W$ luminosity inside the proton and the numerous SM backgrounds that can fake the signal certainly make the measurement harder, but, as a matter of fact, already at the partonic level, the onset of the strong scattering is delayed to higher energies due to a large pollution from the scattering of the transverse polarizations~\cite{SILH2}, as illustrated in Fig.~\ref{fig:WpWpTOWpWp}.

In composite Higgs models, another direct probe of the strong dynamics at the origin of EWSB is the cross section for the double Higgs production. Indeed, the Higgs boson appears as a pseudo Nambu--Goldstone boson and  its properties are directly related to those of the other exact (eaten) Goldstones, corresponding to the longitudinal $W, Z$ gauge bosons. Thus, a generic prediction is that the strong gauge boson scattering is accompanied by strong production of Higgs pairs.The amplitudes for double Higgs production grow with the center-of-mass energy as
\begin{equation}
{\cal A}\left( Z^0_L Z^0_L \to hh \right) =
{\cal A}\left( W^+_L W^-_L \to hh \right)=\frac{c_H s}{f^2}.
\label{eq:doubleHiggs}
\end{equation}
Therefore a significant enhancement over the (negligible) SM rate for the production of two Higgs bosons at high $p_T$, along  with two forward jets associated with the two primary partons that radiated the $W_LW_L$ pair, is expected. An explorative analysis~\cite{SILH2} has shown that the best channel for discovery involves 3 leptons in the final states, with both Higgs bosons decaying to $W^+ W^-$: $pp \to hhjj \to 4Wjj \to l^+l^-l^\pm   \, {E\hspace{-.23cm} \big/ {}_\textrm{T}} \,4j$. The final states are undeniably more complicated than in the analyses of gauge boson scattering and come with smaller branching ratios, but at least the double Higgs production does not suffer from pollution from the transverse modes and it is the only process that gives access to the quadratic coupling $b$ of the Lagrangian~(\ref{eq:scalar}) and allows to test its relation to the linear coupling, $a$, as predicted by the structure of the higher dimension operators~(\ref{eq:silh}): $a=1-c_H v^2/(2f^2), b=1-2 c_H v^2/f^2$.
A Monte-Carlo simulation with simple kinematic cuts concludes~\cite{SILH2} that the signal significance at the LHC operating at $\sqrt{s}=14$~TeV with 300~fb$^{-1}$ will be limited to about 2.5 standard deviations for $v^2/f^2=0.8$. With an upgrade of the LHC luminosity (sLHC program), a 5$\sigma$ discovery can be reached with less than 1~ab$^{-1}$ of integrated luminosity.

While the effective Lagrangian~\ref{eq:silh} elegantly captures the LHC physics of composite Higgs models up to a scale of the order of 10~TeV,  explicit holographic models constructed in 5D warped space provide a valid description in the far UV up to the energies close to the Planck scale and give a new and interesting twist to the question of gauge coupling unification. Not only the running of the gauge couplings receives a contribution from all the resonances of the strong sector (the KK states in the 5D picture) but it also loses the contribution of the Higgs (and the top) above the weak scale. And an appealing unification seems to follow from this minimal set-up with a degree of accuracy comparable to the one reached in the MSSM~\cite{Agashe:2005vg}.

While my presentation has been focussed on the gauge sector, the fermionic sector of composite Higgs models, in particular in the top sector, provides also very interesting signatures easily accessible at the first stages of the LHC operation~\cite{Anastasiou:2009rv, Gillioz:2008hs, Carena:2007tn, Contino:2008hi, Lodone:2008yy, Pomarol:2008bh,  deSandes:2008yx, Mrazek:2009yu}. In particular the same-sign dilepton final states offer a sensitive probe to the top partners~\cite{Contino:2008hi, Mrazek:2009yu} with a discovery potential up to 500~GeV (resp. 1~TeV) with about 50~pb$^{-1}$ (resp. 15~fb$^{-1}$).

In conclusion, in the plausible situation that the LHC sees a Higgs boson and no other direct evidence of new physics, it will not be immediate to determine the true nature of this Higgs boson and tell for sure if it is an elementary particle or a composite bound state emerging from a strongly interacting sector. In that situation, a physics case for a linear collider~\cite{Djouadi:2007ik} together with the sLHC~\cite{Gianotti:2002xx}  can be easily made. 


\section{5D Higgsless models}
\label{sec:Higgsless}


In composite models, when the compositeness scale gets close to the weak scale, the Higgs boson effectively decouples. This Higgsless limit is easily reached in 5D dimensional setups and offers a new point of view on the mass problem. In a sense, the EWSB itself is achieved via boundary conditions (rather than by a Higgs vacuum expectation value). According to the Einstein's relation between the mass and the momentum ($\vec{p}_3$ denotes the momentum along the usual 3 spatial dimension and $p_\perp$ is the momentum along the extra dimension):
\begin{equation}
m^2=E^2-\vec{p}^2_3-p^2_\perp,
\end{equation}
a transverse momentum, $p_\perp$,  simply appears as a mass from the 4D point of view. And the mass problem reduces to a problem of quantum mechanics in a box: suitable boundary conditions will generate a transverse momentum for the appropriate particles. Nonetheless, an immediate question arises: is it better to generate a transverse momentum than to introduce by hand a symmetry breaking mass for the gauge field? In other words, how is unitarity restored? In full generality, the elastic scattering amplitude of a massive Kaluza--Klein (KK) gauge field would have terms that grow with the fourth and the second powers of the energy
\begin{equation}
\mathcal{A} = \mathcal{A}^{(4)} \left( \frac{E}{M}\right)^4 + \mathcal{A}^{(2)} \left( \frac{E}{M}\right)^2 + \mathcal{A}^{(0)} + \ldots 
\end{equation}
In the SM, $\mathcal{A}^{(4)}$ is automatically vanishing due to gauge invariance, while $\mathcal{A}^{(2)}$ vanishes via the exchange of the physical Higgs boson. In 5D Higgsless models, the unitarization of the $WW$ scattering amplitudes follows from the exchange of all the KK excitations of the $W$. In order for this unitarization to actually happen, the couplings and the masses of the KK excitations have to obey the following sum rules\footnote{$g^2_{WWWW}$ is the quartic $W$ self-coupling, $g_{WWX}$ is the cubic coupling between two $W$'s and $X$ and $Z^{(n)}$ denote the KK excitations of the $Z$ . The two sum rules~(\ref{eq:sumrules}) correspond to the $W^\pm W^\pm \to W^\pm W^\pm$ channel and similar sum rules will apply to other $WZ$ channels.}~\cite{Csaki:2003dt}:
\begin{eqnarray}
\label{eq:sumrules}
\begin{gathered}
g^2_{WWWW} = g^2_{WW\gamma} + g^2_{WWZ}+\sum_n g^2_{WWZ^{(n)}}\\
4 g^2_{WWWW}  M_W^2= 3 g^2_{WWZ} M_Z^2+ 3 \sum_n g^2_{WWZ^{(n)}} M^2_{Z^{(n)}}.
\end{gathered}
\end{eqnarray}
The effective couplings among the KK states are dictated by the gauge structure  of the 5D theory and it is easy to show that the two sum rules are automatically satisfied, provided there is no hard breaking of gauge invariance, i.e., if the 5D gauge fields obey Dirichlet or Neumann boundary conditions at the end points of the 5${}^\textrm{\tiny th}$ dimension.

Is this KK unitarization a counter example of the theorem by Cornwall et al. mentioned in Section~\ref{sec:strongEWSB}? Not exactly since, as the energy grows, more and more inelastic channels open up  and they ultimately saturate the perturbative unitarity bound~\cite{Papucci:2004ip, SekharChivukula:2001hz} around the scale $4\pi^4 M_W^2/(g^2 M_{Z^{(1)}})$ ($g$ is the usual $SU(2)$ gauge coupling). Thus, we can see that for this scale to be substantially above the usual unitarity violation scale 
of the SM without a Higgs, one needs to have the first KK resonance to be as light as possible. The existence of a $W'$ and a $Z'$ below 1~TeV and with significant cubic couplings to the SM gauge bosons is a robust prediction of Higgsless models.

\begin{figure}[htbp]
\begin{center}
\includegraphics[width=.75\textwidth]{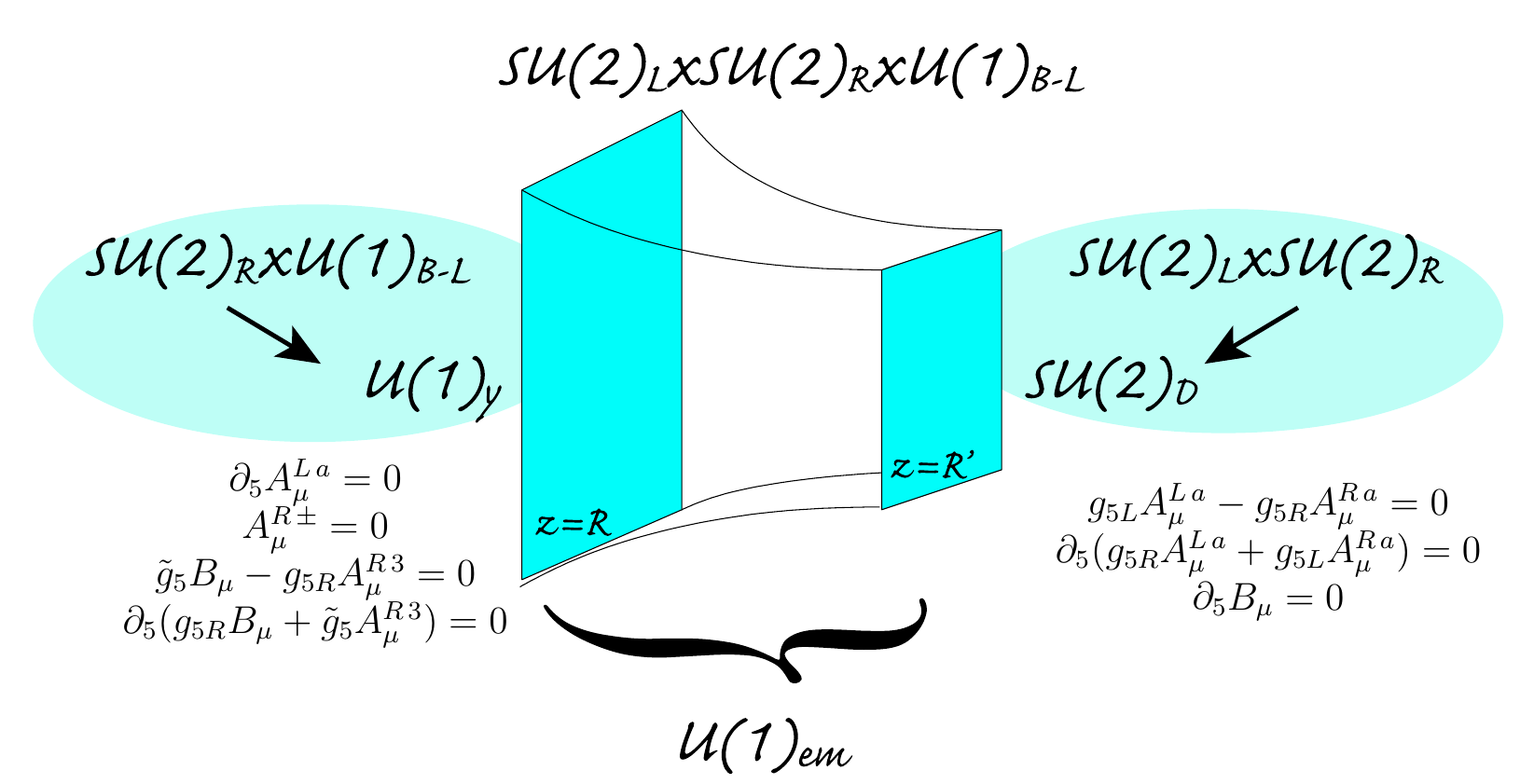}
\caption[]{The symmetry-breaking structure of the warped Higgsless
model~\cite{Csaki:2003zu}. The model considers a 5D gauge theory in a fixed gravitational anti-de-Sitter (AdS) background.
The UV~brane (sometimes called the Planck brane) is located at $z=R$ and the IR~brane (also called the TeV brane) is located at $z=R'$. $R$ is the AdS curvature scale. In conformal coordinates, the AdS metric is given by
$
ds^2=  \left( R/z \right)^2   \Big( \eta_{\mu \nu} dx^\mu dx^\nu - dz^2 \Big).
$
}
\label{fig:higgsless}
\end{center}
\end{figure}

For a concrete implementation of the Higgsless idea, one can conveniently use a warped extra dimension~\cite{Csaki:2003zu}\footnote{Alternatively, deconstruction versions have been proposed (see for instance Refs.~\cite{Belyaev:2009ve,Accomando:2008dm} and references therein), which extend the pioneering BESS model~\cite{Casalbuoni:1985kq}}, which (i)~automatically ensures a mass gap between the $W, Z$ and their KK excitations and (ii)~is instrumental to enforce a custodial $SU(2)$ symmetry that is implemented as a bulk $SU(2)_L\times SU(2)_R \times U(1)_{B-L}$ gauge symmetry. The way the proper symmetry breaking is achieved is, see Fig.~\ref{fig:higgsless}, by breaking the bulk gauge group down to the SM group $SU(2)_L \times U(1)_Y$ on the UV brane, thus ensuring that the additional gauge symmetry only manifests itself as a global symmetry among the KK states. The electroweak symmetry breaking is then achieved by breaking $SU(2)_L \times SU(2)_R$ to $SU(2)_D$ on the IR brane, which, according to the AdS/CFT principles, appears as a spontaneous breaking of the effective theory.

In the SM, the physical Higgs boson is instrumental in bringing the theory in good agreement with EW data. In the 5D Higgsless model described above, the first corrections to the EW precision observables are given by~\cite{Barbieri:2003pr, Cacciapaglia:2004jz}
\begin{equation}
S \equiv \frac{6\pi}{g^2 \log{R'/R}}\sim 1.15, \ \ T\equiv 0.
\end{equation}
Thus, while the $T$ parameter is protected by the built-in custodial symmetry, the $S$ parameter is too large. Nonetheless, $S$ can be tuned away by delocalizing the (light) fermions in the bulk~\cite{Cacciapaglia:2004rb, Foadi:2004ps}. For a particular profile of the fermion wavefunctions, the fermions will almost decouple from the gauge boson KK excitations and the resulting $S$ parameter will vanish. However, some doubts about the radiative stability of this delocalization setup have been raised~\cite{Dawson:2008as}. Furthermore, delocalizing the fermions in the bulk also precludes a nice dynamical origin of the fermion mass hierarchy and the model has to be supplemented with an additional flavor structure (see for instance Ref.~\cite{Cacciapaglia:2007fw}).

The remaining problem with precision EW data is the compatibility of a large 
top quark mass with the observed $Zb_L\bar{b}_L$ coupling. The large top mass requires both the 
left-handed and right-handed top quarks to have a profile which is localized toward the IR brane. However, if the left-handed top (and thus the left-handed bottom) are too close to the IR brane, then the gauge couplings of the bottom will be too different from the down and strange quarks. This problem may be avoided by separating the physics which generates the top quark mass~\cite{Cacciapaglia:2005pa},  by allowing a Higgs vacuum expectation value to extend slightly into the bulk~\cite{Cacciapaglia:2006mz} or by introducing new fermionic states with exotic charges~\cite{Agashe:2006at, Cacciapaglia:2006gp}.

The most distinctive feature of the Higgsless models is, of course, the absence of a physical scalar 
state in the spectrum. Yet, {\it the absence of proof is not the proof of the absence}  and some other models exist in which the Higgs is unobservable at the LHC (for a recent review, see for instance Ref.~\cite{DeRoeck:2009id}). Fortunately, Higgsless models are characterized by other distinctive features, such as (i)~the presence of spin-1 KK resonances with the $W, Z$ quantum numbers, (ii)~some slight deviations in the universality of the light fermion couplings to the SM gauge bosons and (iii)~some deviations in the gauge boson self-interactions compared with the SM. References~\cite{Birkedal:2004au, Englert:2008wp} studied the production of the lightest KK excitations of the $W$ and the $Z$ via vector boson fusion. The most recent study~\cite{Martin:2009gi} included also the more model-dependent possibility of Drell--Yan production. 
At the LHC, about 10~fb$^{-1}$ of luminosity will be necessary for the discovery of the resonances in the 700~GeV mass range. A precise measurement of the couplings of these resonances or the search for some deviations in the SM couplings will require a more precise machine, such as an ILC or CLIC.


\section{Conclusion}

The SM has emerged as a successful description, at the quantum level, of the interactions among the elementary particles  but it is at odds in what concerns the masses of these elementary particles.
EW interactions certainly need  three Nambu--Goldstone bosons to provide a mass to the $W^\pm$ and the $Z$ gauge bosons. But they also need new dynamics to act as a UV moderator and ensure a proper decoupling at high energy of the extra polarizations associated to the mass of these spin-1 fields. After many years of theoretical speculations and in the absence of any direct experimental evidences, it is not yet possible to tell whether the strength of this new dynamics is weak or strong. In many regards, this question is equivalent to asking whether a light Higgs boson exists or not. However it is also possible and plausible that a light and narrow Higgs-like scalar does exist but it is actually a bound state from some strong dynamics not much above the weak scale. Such composite models provide a continuous dynamical deformation of the SM, with the same spectrum as the SM up to 2$\div$3~TeV.

The LHC is prepared to discover the Higgs boson or whatever replaces it. To this end, the collaboration between experimentalists and theorists is more important than ever to make sure, for instance, that no unexpected physics is missed because of triggers and cuts. In this regards, signature-motivated approaches like `unparticles'~\cite{Georgi:2007ek}, `hidden valleys'~\cite{Strassler:2006im} or `quirks'~\cite{Kang:2008ea} should be encouraged.

Also, it should not be forgotten that the LHC will be a top(-quark) machine.  And there are many reasons to believe that the top quark can be an important agent in the dynamics triggering the electroweak symmetry breaking. 

Finally, it is also worth mentioning that most theories for the Fermi scale can be probed outside of colliders.
The numerous experiments searching for Dark Matter nicely exemplify the strong cosmo-astro particle connection, and, for sure, if weakly interacting massive particles (WIMPs) are part of the dynamics of EWSB, their direct or indirect detections would provide valuable information. An intriguing signature~\cite{HiggsSpace} arises when the WIMPS can annihilate directly into a photon and a Higgs boson, giving rise to a line in the gamma ray spectrum whose position reflects the Higgs mass: the observation of such a line would be the first direct observation of a Higgs production process! The complementarity between astrophysics and collider physics is not restricted to Dark Matter. Another compelling example 
concerns gravitational waves: a background of stochastic gravitational waves peaked around  mHz frequencies would be an indication of a strong EW phase transition due for instance to enhanced Higgs self-couplings~\cite{Grojean:2004xa, Grojean:2006bp}.

In any case, more than ever, experimental data are eagerly awaited to disentangle what may be the most pressing question faced by particle physics today: How is electroweak symmetry broken?

\section*{Acknowledgements}
It is pleasure to thank my various collaborators, especially C.~Cs\'aki and J.~Terning for fruitful works on Higgsless projects and G.~Giudice, A.~Pomarol and R. Rattazzi for our construction of the effective description of composite Higgs models. A lot of material presented during this talk is based on two on-going projects with R.~Contino, M.~Moretti, F.~Piccinini and R.~Rattazzi and with J.R.~Espinosa and 
M.~M\"uhlleitner and these words are to tell them all my gratitude for their support. 
I thank R.~Contino, J.R.~Espinosa, G.~ Giudice, M.~M\"uhlleitner, S.~Pokorski and A.~Weiler for useful comments on the manuscript.  
I also thank the organizers of the EPS-HEPP'09 conference and the board of the EPS-HEPP division for inviting me to present this review talk. My work has been supported in part by the `MassTeV' ERC advanced 
grant 226371.


\begin{thebibliography}{99}
 
\bibitem{Nambu:1960xd}
  Y.~Nambu,
  Phys.\ Rev.\ Lett.\  {\bf 4} (1960) 380.
  
\bibitem{Goldstone:1961eq}
  J.~Goldstone,
  Nuovo Cim.\  {\bf 19} (1961) 154.

\bibitem{Englert:1964et}
  F.~Englert and R.~Brout,
  Phys.\ Rev.\ Lett.\  {\bf 13} (1964) 321.
  
\bibitem{Higgs:1964pj}
  P.~W.~Higgs,
  Phys.\ Rev.\ Lett.\  {\bf 13} (1964) 508.
      
\bibitem{Barate:2003sz}
  R.~Barate {\it et al.}  [LEP Working Group for Higgs boson searches],
  Phys.\ Lett.\  B {\bf 565} (2003) 61
  [arXiv:hep-ex/0306033].

\bibitem{Amsler:2008zzb}
  C.~Amsler {\it et al.}  [Particle Data Group],
  Phys.\ Lett.\  B {\bf 667} (2008) 1.

 \bibitem{Hoecker:2009gd}
  A.~Hoecker, these conference proceedings,
  arXiv:0909.0961 [hep-ph].
  
\bibitem{Djouadi:2005gi}
  A.~Djouadi,
  Phys.\ Rept.\  {\bf 457} (2008) 1
  [arXiv:hep-ph/0503172].
  
\bibitem{Weisskopf:1939}
V.F.~Weisskopf,
Phys.\ Rev. {\bf 56}, 72 (1939).
  
\bibitem{'tHooft:1979bh}
  G.~'t Hooft,
  NATO Adv.\ Study Inst.\ Ser.\ B Phys.\  {\bf 59} (1980) 135.

\bibitem{Veltman:1980mj}
  M.~J.~G.~Veltman,
  Acta Phys.\ Polon.\  B {\bf 12} (1981) 437.

\bibitem{Wilson:1973jj}
  K.~G.~Wilson and J.~B.~Kogut,
  Phys.\ Rept.\  {\bf 12} (1974) 75.
  
\bibitem{Luscher:1987ek}
  M.~Luscher and P.~Weisz,
  Nucl.\ Phys.\  B {\bf 295} (1988) 65.



\bibitem{Linde:1975sw}
  A.~D.~Linde,
  JETP Lett.\  {\bf 23} (1976) 64
  [Pisma Zh.\ Eksp.\ Teor.\ Fiz.\  {\bf 23} (1976) 73].
  
\bibitem{Weinberg:1976pe}
  S.~Weinberg,
  Phys.\ Rev.\ Lett.\  {\bf 36} (1976) 294.

\bibitem{Manton:1979kb}
  N.~S.~Manton,
  Nucl.\ Phys.\  B {\bf 158} (1979) 141.

  
\bibitem{Hosotani:1983xw}
  Y.~Hosotani,
  Phys.\ Lett.\  B {\bf 126} (1983) 309.

\bibitem{Antoniadis:2001cv}
  I.~Antoniadis, K.~Benakli and M.~Quiros,
  New J.\ Phys.\  {\bf 3} (2001) 20
  [arXiv:hep-th/0108005].

\bibitem{Csaki:2002ur}
  C.~Csaki, C.~Grojean and H.~Murayama,
  Phys.\ Rev.\  D {\bf 67} (2003) 085012
  [arXiv:hep-ph/0210133].

\bibitem{Serone:2009kf}
  M.~Serone,
  arXiv:0909.5619 [hep-ph].

\bibitem{Bezrukov:2007ep}
  F.~L.~Bezrukov and M.~Shaposhnikov,
  Phys.\ Lett.\  B {\bf 659} (2008) 703
  [arXiv:0710.3755 [hep-th]].

\bibitem{Grojean:2007zz}
  C.~Grojean,
  Phys.\ Usp.\  {\bf 50} (2007) 1
  [Usp.\ Fiz.\ Nauk {\bf 177} (2007) 3].

\bibitem{Bellazzini:2009xt}
  B.~Bellazzini, C.~Csaki, A.~Falkowski and A.~Weiler,
  arXiv:0906.3026 [hep-ph].

\bibitem{Bellazzini:2009kw}
  B.~Bellazzini, C.~Csaki, A.~Falkowski and A.~Weiler,
  arXiv:0910.3210 [hep-ph].
  
\bibitem{Georgi:1981gk}
  H.~Georgi and I.~N.~McArthur,
  Nucl.\ Phys.\  B {\bf 202} (1982) 382.
  
\bibitem{Harnik:2003rs}
  R.~Harnik, G.~D.~Kribs, D.~T.~Larson and H.~Murayama,
  Phys.\ Rev.\  D {\bf 70} (2004) 015002
  [arXiv:hep-ph/0311349].

\bibitem{Diaz:1994pk}
  M.~A.~Diaz and T.~J.~Weiler,
  arXiv:hep-ph/9401259.

\bibitem{Hall:2001zb}
  L.~J.~Hall, Y.~Nomura and D.~Tucker-Smith,
  Nucl.\ Phys.\  B {\bf 639} (2002) 307
  [arXiv:hep-ph/0107331].
  
\bibitem{Cacciapaglia:2006mz}
  G.~Cacciapaglia, C.~Csaki, G.~Marandella and J.~Terning,
  JHEP {\bf 0702} (2007) 036
  [arXiv:hep-ph/0611358].
   
\bibitem{Contino:2007zz}
  R.~Contino and A.~Pomarol,
  Comptes Rendus Physique {\bf 8} (2007) 1058.

\bibitem{Katz:2005au}
  E.~Katz, A.~E.~Nelson and D.~G.~E.~Walker,
  JHEP {\bf 0508} (2005) 074
  [arXiv:hep-ph/0504252].
  
\bibitem{Su:2009fz}
  S.~Su and B.~Thomas,
  Phys.\ Rev.\  D {\bf 79} (2009) 095014
  [arXiv:0903.0667 [hep-ph]].

\bibitem{ArkaniHamed:2002qx}
  N.~Arkani-Hamed, A.~G.~Cohen, E.~Katz, A.~E.~Nelson, T.~Gregoire and J.~G.~Wacker,
  JHEP {\bf 0208} (2002) 021
  [arXiv:hep-ph/0206020].

\bibitem{ArkaniHamed:2002qy}
  N.~Arkani-Hamed, A.~G.~Cohen, E.~Katz and A.~E.~Nelson,
  JHEP {\bf 0207} (2002) 034
  [arXiv:hep-ph/0206021].
  
\bibitem{Hsieh:2008jg}
  K.~Hsieh and C.~P.~Yuan,
  Phys.\ Rev.\  D {\bf 78} (2008) 053006
  [arXiv:0806.2608 [hep-ph]].
  
\bibitem{Delgado:2008px}
  A.~Delgado, J.~R.~Espinosa, J.~M.~No and M.~Quiros,
  JHEP {\bf 0811} (2008) 071
  [arXiv:0804.4574 [hep-ph]].

\bibitem{Patt:2006fw}
  B.~Patt and F.~Wilczek,
  arXiv:hep-ph/0605188.

\bibitem{Porto:2007ed}
  R.~A.~Porto and A.~Zee,
  Phys.\ Lett.\  B {\bf 666} (2008) 491
  [arXiv:0712.0448 [hep-ph]].

\bibitem{Wudka:1988az}
  J.~Wudka,
17th Int. Colloq. on Group Theoretical Methods in Physics, Sainte-Adele, Canada, 1988.  
  
  \bibitem{Schmaltz:2004de}
  M.~Schmaltz,
  JHEP {\bf 0408} (2004) 056
  [arXiv:hep-ph/0407143].

\bibitem{Lee:2009up}
  J.~S.~Lee, Y.~Peters, A.~Pilaftsis and C.~Schwanenberger,
  arXiv:0909.1749 [hep-ph].

\bibitem{Chacko:2005pe}
  Z.~Chacko, H.~S.~Goh and R.~Harnik,
  Phys.\ Rev.\ Lett.\  {\bf 96} (2006) 231802
  [arXiv:hep-ph/0506256].
  
\bibitem{Stancato:2008mp}
  D.~Stancato and J.~Terning,
  arXiv:0807.3961 [hep-ph].

 \bibitem{Dermisek:2009si}
  R.~Dermisek,
  Mod.\ Phys.\ Lett.\  A {\bf 24} (2009) 1631
  [arXiv:0907.0297 [hep-ph]].

\bibitem{Heinemeyer:2006px}
  S.~Heinemeyer, W.~Hollik, D.~Stockinger, A.~M.~Weber and G.~Weiglein,
  JHEP {\bf 0608} (2006) 052
  [arXiv:hep-ph/0604147].

\bibitem{Giudice:2006sn}
  G.~F.~Giudice and R.~Rattazzi,
  Nucl.\ Phys.\  B {\bf 757} (2006) 19
  [arXiv:hep-ph/0606105].
  
\bibitem{Casas:2003jx}
  J.~A.~Casas, J.~R.~Espinosa and I.~Hidalgo,
  JHEP {\bf 0401} (2004) 008
  [arXiv:hep-ph/0310137].
  
\bibitem{Fayet:1975yi}
  P.~Fayet,
  Nucl.\ Phys.\  B {\bf 113} (1976) 135.

\bibitem{Batra:2003nj}
  P.~Batra, A.~Delgado, D.~E.~Kaplan and T.~M.~P.~Tait,
  JHEP {\bf 0402} (2004) 043
  [arXiv:hep-ph/0309149].
    
\bibitem{Birkedal:2004xi}
  A.~Birkedal, Z.~Chacko and M.~K.~Gaillard,
  JHEP {\bf 0410} (2004) 036
  [arXiv:hep-ph/0404197].
 
\bibitem{Chankowski:2004mq}
  P.~H.~Chankowski, A.~Falkowski, S.~Pokorski and J.~Wagner,
  Phys.\ Lett.\  B {\bf 598} (2004) 252
  [arXiv:hep-ph/0407242].
  
\bibitem{Brignole:2003cm}
  A.~Brignole, J.~A.~Casas, J.~R.~Espinosa and I.~Navarro,
  Nucl.\ Phys.\  B {\bf 666} (2003) 105
  [arXiv:hep-ph/0301121].
  
\bibitem{Dine:2007xi}
  M.~Dine, N.~Seiberg and S.~Thomas,
  Phys.\ Rev.\  D {\bf 76} (2007) 095004
  [arXiv:0707.0005 [hep-ph]].
 
\bibitem{Gherghetta:1995dv}
  T.~Gherghetta, C.~F.~Kolda and S.~P.~Martin,
  Nucl.\ Phys.\  B {\bf 468} (1996) 37
  [arXiv:hep-ph/9510370].
  
\bibitem{Antoniadis:2007xc}
  I.~Antoniadis, E.~Dudas and D.~M.~Ghilencea,
  JHEP {\bf 0803} (2008) 045
  [arXiv:0708.0383 [hep-th]].
  
\bibitem{Antoniadis:2008es}
  I.~Antoniadis, E.~Dudas, D.~M.~Ghilencea and P.~Tziveloglou,
  Nucl.\ Phys.\  B {\bf 808} (2009) 155
  [arXiv:0806.3778 [hep-ph]].
  
\bibitem{Strumia:1999jm}
  A.~Strumia,
  Phys.\ Lett.\  B {\bf 466} (1999) 107
  [arXiv:hep-ph/9906266].
  
\bibitem{Blum:2009na}
  K.~Blum, C.~Delaunay and Y.~Hochberg,
  arXiv:0905.1701 [hep-ph].

\bibitem{Blum:2008ym}
  K.~Blum and Y.~Nir,
  Phys.\ Rev.\  D {\bf 78} (2008) 035005
  [arXiv:0805.0097 [hep-ph]].
  
\bibitem{Bernal:2009hd}
  N.~Bernal, K.~Blum, Y.~Nir and M.~Losada,
  JHEP {\bf 0908} (2009) 053
  [arXiv:0906.4696 [hep-ph]].

\bibitem{Carena:2009gx}
  M.~Carena, K.~Kong, E.~Ponton and J.~Zurita,
  arXiv:0909.5434 [hep-ph].

\bibitem{Cassel:2009ps}
  S.~Cassel, D.~M.~Ghilencea and G.~G.~Ross,
  arXiv:0903.1115 [hep-ph].
  
\bibitem{UrangaEPS}
  A.~Uranga, these conference proceedings.
  
\bibitem{Gherghetta:2006ha}
  T.~Gherghetta,
  arXiv:hep-ph/0601213.

\bibitem{Inami:1992rb}
  T.~Inami, C.~S.~Lim and A.~Yamada,
  Mod.\ Phys.\ Lett.\  A {\bf 7} (1992) 2789.
  
\bibitem{ArkaniHamed:2001nc}
  N.~Arkani-Hamed, A.~G.~Cohen and H.~Georgi,
  Phys.\ Lett.\  B {\bf 513} (2001) 232
  [arXiv:hep-ph/0105239].
  
\bibitem{Grinstein:2008kt}
  B.~Grinstein and M.~Trott,
  JHEP {\bf 0811} (2008) 064
  [arXiv:0808.2814 [hep-ph]].

\bibitem{Casas:2005ev}
  J.~A.~Casas, J.~R.~Espinosa and I.~Hidalgo,
  JHEP {\bf 0503} (2005) 038
  [arXiv:hep-ph/0502066].

\bibitem{Hill:2007zv}
  C.~T.~Hill and R.~J.~Hill,
  Phys.\ Rev.\  D {\bf 76} (2007) 115014
  [arXiv:0705.0697 [hep-ph]].
  
 \bibitem{Csaki:2008se}
  C.~Csaki, J.~Heinonen, M.~Perelstein and C.~Spethmann,
  arXiv:0804.0622 [hep-ph].
  
\bibitem{Cheng:2003ju}
  H.~C.~Cheng and I.~Low,
  JHEP {\bf 0309} (2003) 051
  [arXiv:hep-ph/0308199].
  
\bibitem{Carena:2006jx}
  M.~S.~Carena, J.~Hubisz, M.~Perelstein and P.~Verdier,
  Phys.\ Rev.\  D {\bf 75} (2007) 091701
  [arXiv:hep-ph/0610156].
  
\bibitem{Han:2003wu}
T.~Han, H.~E.~Logan, B.~McElrath and L.~T.~Wang,
  Phys.\ Rev.\  D {\bf 67} (2003) 095004
  [arXiv:hep-ph/0301040].

\bibitem{Perelstein:2003wd}
  M.~Perelstein, M.~E.~Peskin and A.~Pierce,
  Phys.\ Rev.\  D {\bf 69} (2004) 075002
  [arXiv:hep-ph/0310039].
  
\bibitem{Chanowitz:1985hj}
  M.~S.~Chanowitz and M.~K.~Gaillard,
  Nucl.\ Phys.\  B {\bf 261} (1985) 379.
  
\bibitem{SILH2}  
R.~Contino, C.~Grojean, M.~Moretti, F.~Piccinini and R.~Rattazzi, to appear.  
 
\bibitem{Bellazzini:2008zy}
  B.~Bellazzini, S.~Pokorski, V.~S.~Rychkov and A.~Varagnolo,
  JHEP {\bf 0811} (2008) 027
  [arXiv:0805.2107 [hep-ph]].
  
\bibitem{Cornwall:1973tb}
  J.~M.~Cornwall, D.~N.~Levin and G.~Tiktopoulos,
  Phys.\ Rev.\ Lett.\  {\bf 30} (1973) 1268
  [Erratum-ibid.\  {\bf 31} (1973) 572].

  
\bibitem{Agashe:2005vg}
  K.~Agashe, R.~Contino and R.~Sundrum,
  Phys.\ Rev.\ Lett.\  {\bf 95}, 171804 (2005)
  [arXiv:hep-ph/0502222].

\bibitem{Agashe:2004rs}
  K.~Agashe, R.~Contino and A.~Pomarol,
  Nucl.\ Phys.\  B {\bf 719} (2005) 165
  [arXiv:hep-ph/0412089].
  
\bibitem{Low:2009di}
  I.~Low, R.~Rattazzi and A.~Vichi,
  arXiv:0907.5413 [hep-ph].
  
\bibitem{Hill:2002ap}
  C.~T.~Hill and E.~H.~Simmons,
  Phys.\ Rept.\  {\bf 381} (2003) 235
  [Erratum-ibid.\  {\bf 390} (2004) 553]
  [arXiv:hep-ph/0203079].

\bibitem{Sannino:2008kg}
  F.~Sannino,
  arXiv:0806.3575 [hep-ph].

\bibitem{Gripaios:2009pe}
  B.~Gripaios, A.~Pomarol, F.~Riva and J.~Serra,
  JHEP {\bf 0904} (2009) 070
  [arXiv:0902.1483 [hep-ph]].

\bibitem{Georgi:1984af}
  H.~Georgi and D.~B.~Kaplan,
  Phys.\ Lett.\  B {\bf 145} (1984) 216.

\bibitem{Randall:1999ee}
  L.~Randall and R.~Sundrum,
  Phys.\ Rev.\ Lett.\  {\bf 83} (1999) 3370
  [arXiv:hep-ph/9905221].

\bibitem{ArkaniHamed:2000ds}
  N.~Arkani-Hamed, M.~Porrati and L.~Randall,
  JHEP {\bf 0108} (2001) 017
  [arXiv:hep-th/0012148].
  
  \bibitem{Rattazzi:2000hs}
  R.~Rattazzi and A.~Zaffaroni,
  JHEP {\bf 0104} (2001) 021
  [arXiv:hep-th/0012248].

\bibitem{Giudice:2007fh}
  G.~F.~Giudice, C.~Grojean, A.~Pomarol and R.~Rattazzi,
  JHEP {\bf 0706} (2007) 045
  [arXiv:hep-ph/0703164].

\bibitem{Barbieri:2007bh}
  R.~Barbieri, B.~Bellazzini, V.~S.~Rychkov and A.~Varagnolo,
  Phys.\ Rev.\  D {\bf 76} (2007) 115008
  [arXiv:0706.0432 [hep-ph]].

\bibitem{Anastasiou:2009rv}
  C.~Anastasiou, E.~Furlan and J.~Santiago,
  arXiv:0901.2117 [hep-ph].

\bibitem{Contino:2009ez}
  R.~Contino,
  arXiv:0908.3578 [hep-ph].
  
\bibitem{Agashe:2005dk}
  K.~Agashe and R.~Contino,
  Nucl.\ Phys.\  B {\bf 742} (2006) 59
  [arXiv:hep-ph/0510164].

\bibitem{Carena:2007ua}
  M.~S.~Carena, E.~Ponton, J.~Santiago and C.~E.~M.~Wagner,
  Phys.\ Rev.\  D {\bf 76} (2007) 035006
  [arXiv:hep-ph/0701055].

\bibitem{Gillioz:2008hs}
  M.~Gillioz,
  Phys.\ Rev.\  D {\bf 80} (2009) 055003
  [arXiv:0806.3450 [hep-ph]].
  
\bibitem{Bouchart:2008vp}
  C.~Bouchart and G.~Moreau,
  Nucl.\ Phys.\  B {\bf 810} (2009) 66
  [arXiv:0807.4461 [hep-ph]].
  
\bibitem{Contino:2006qr}
  R.~Contino, L.~Da Rold and A.~Pomarol,
  Phys.\ Rev.\  D {\bf 75} (2007) 055014
  [arXiv:hep-ph/0612048].
  
\bibitem{D'Ambrosio:2002ex}
  G.~D'Ambrosio, G.~F.~Giudice, G.~Isidori and A.~Strumia,
  Nucl.\ Phys.\  B {\bf 645} (2002) 155
  [arXiv:hep-ph/0207036].
    
\bibitem{Agashe:2009di}
  K.~Agashe and R.~Contino,
  arXiv:0906.1542 [hep-ph].
  
\bibitem{Azatov:2009na}
  A.~Azatov, M.~Toharia and L.~Zhu,
  Phys.\ Rev.\  D {\bf 80} (2009) 035016
  [arXiv:0906.1990 [hep-ph]].

\bibitem{ArkaniHamed:1999dc}
  N.~Arkani-Hamed and M.~Schmaltz,
  Phys.\ Rev.\  D {\bf 61} (2000) 033005
  [arXiv:hep-ph/9903417].

\bibitem{Grossman:1999ra}
  Y.~Grossman and M.~Neubert,
  Phys.\ Lett.\  B {\bf 474} (2000) 361
  [arXiv:hep-ph/9912408].

\bibitem{Kaplan:1991dc}
  D.~B.~Kaplan,
  Nucl.\ Phys.\  B {\bf 365} (1991) 259.

\bibitem{Huber:2003tu}
  S.~J.~Huber,
  Nucl.\ Phys.\  B {\bf 666} (2003) 269
  [arXiv:hep-ph/0303183].

\bibitem{Agashe:2004ay}
  K.~Agashe, G.~Perez and A.~Soni,
  Phys.\ Rev.\ Lett.\  {\bf 93} (2004) 201804
  [arXiv:hep-ph/0406101].

\bibitem{Csaki:2008zd}
  C.~Csaki, A.~Falkowski and A.~Weiler,
  JHEP {\bf 0809} (2008) 008
  [arXiv:0804.1954 [hep-ph]].

\bibitem{Agashe:2008uz}
  K.~Agashe, A.~Azatov and L.~Zhu,
  Phys.\ Rev.\  D {\bf 79} (2009) 056006
  [arXiv:0810.1016 [hep-ph]].
  
\bibitem{Buras:2009if}
  A.~J.~Buras,
 these conference proceedings,
  arXiv:0910.1032 [hep-ph].
    
\bibitem{Djouadi:2007fm}
  A.~Djouadi and G.~Moreau,
  Phys.\ Lett.\  B {\bf 660} (2008) 67
  [arXiv:0707.3800 [hep-ph]].
  
\bibitem{Falkowski:2007hz}
  A.~Falkowski,
  Phys.\ Rev.\  D {\bf 77} (2008) 055018
  [arXiv:0711.0828 [hep-ph]].

\bibitem{Ball:2007zza}
  G.~L.~Bayatian {\it et al.}  [CMS Collaboration],
  ``CMS technical design report, volume II: Physics performance,''
  J.\ Phys.\ G {\bf 34} (2007) 995.

\bibitem{SILH3}  
J.R.~Espinosa, C.~Grojean and M.~Muehlleitner, to appear.

\bibitem{Manohar:2006gz}
  A.~V.~Manohar and M.~B.~Wise,
  Phys.\ Lett.\  B {\bf 636} (2006) 107
  [arXiv:hep-ph/0601212].
  
\bibitem{Pierce:2006dh}
  A.~Pierce, J.~Thaler and L.~T.~Wang,
  JHEP {\bf 0705} (2007) 070
  [arXiv:hep-ph/0609049].

\bibitem{Duhrssen:2003}
M.~Duhrssen, ATL--PHYS--2003--030.
  
\bibitem{Duhrssen:2004cv}
  M.~Duhrssen, S.~Heinemeyer, H.~Logan, D.~Rainwater, G.~Weiglein and D.~Zeppenfeld,
  Phys.\ Rev.\  D {\bf 70} (2004) 113009
  [arXiv:hep-ph/0406323].

\bibitem{Gianotti:2002xx}
  F.~Gianotti {\it et al.},
  Eur.\ Phys.\ J.\  C {\bf 39} (2005) 293
  [arXiv:hep-ph/0204087].


\bibitem{Djouadi:2007ik}
  G.~Aarons {\it et al.}  [ILC Collaboration],
  arXiv:0709.1893 [hep-ph].
  
\bibitem{AguilarSaavedra:2001rg}
  J.~A.~Aguilar-Saavedra {\it et al.}  [ECFA/DESY LC Physics Working Group],
  arXiv:hep-ph/0106315.
        
\bibitem{Barger:2003rs}
  V.~Barger, T.~Han, P.~Langacker, B.~McElrath and P.~Zerwas,
  Phys.\ Rev.\  D {\bf 67}, 115001 (2003)
  [arXiv:hep-ph/0301097].
   
\bibitem{Accomando:2004sz}
  E.~Accomando {\it et al.}  [CLIC Physics Working Group],
  arXiv:hep-ph/0412251.
      
   
\bibitem{Falkowski:2007iv}
  A.~Falkowski, S.~Pokorski and J.~P.~Roberts,
  JHEP {\bf 0712} (2007) 063
  [arXiv:0705.4653 [hep-ph]].
    
\bibitem{Bagger:1995mk}
  J.~Bagger {\it et al.},
  Phys.\ Rev.\  D {\bf 52} (1995) 3878
  [arXiv:hep-ph/9504426].
    
\bibitem{Butterworth:2002tt}
  J.~M.~Butterworth, B.~E.~Cox and J.~R.~Forshaw,
  Phys.\ Rev.\  D {\bf 65} (2002) 096014
  [arXiv:hep-ph/0201098].
    
 \bibitem{Ballestrero:2009vw}
  A.~Ballestrero, G.~Bevilacqua, D.~B.~Franzosi and E.~Maina,
  arXiv:0909.3838 [hep-ph].
    
\bibitem{Carena:2007tn}
  M.~Carena, A.~D.~Medina, B.~Panes, N.~R.~Shah and C.~E.~M.~Wagner,
  Phys.\ Rev.\  D {\bf 77} (2008) 076003
  [arXiv:0712.0095 [hep-ph]].

\bibitem{Contino:2008hi}
  R.~Contino and G.~Servant,
  JHEP {\bf 0806} (2008) 026
  [arXiv:0801.1679 [hep-ph]].

 \bibitem{Lodone:2008yy}
  P.~Lodone,
  JHEP {\bf 0812} (2008) 029
  [arXiv:0806.1472 [hep-ph]].
  
\bibitem{Pomarol:2008bh}
  A.~Pomarol and J.~Serra,
  Phys.\ Rev.\  D {\bf 78} (2008) 074026
  [arXiv:0806.3247 [hep-ph]].
  
\bibitem{deSandes:2008yx}
  H.~de Sandes and R.~Rosenfeld,
  J.\ Phys.\ G {\bf 36} (2009) 085001
  [arXiv:0811.0984 [hep-ph]].

\bibitem{Mrazek:2009yu}
  J.~Mrazek and A.~Wulzer,
  arXiv:0909.3977 [Unknown].
  
\bibitem{Csaki:2003dt}
  C.~Csaki, C.~Grojean, H.~Murayama, L.~Pilo and J.~Terning,
  Phys.\ Rev.\  D {\bf 69} (2004) 055006
  [arXiv:hep-ph/0305237].

\bibitem{Papucci:2004ip}
  M.~Papucci,
  arXiv:hep-ph/0408058.
  
\bibitem{SekharChivukula:2001hz}
  R.~S.~Chivukula, D.~A.~Dicus and H.~J.~He,
  Phys.\ Lett.\  B {\bf 525} (2002) 175
  [arXiv:hep-ph/0111016].
    
\bibitem{Csaki:2003zu}
  C.~Csaki, C.~Grojean, L.~Pilo and J.~Terning,
  Phys.\ Rev.\ Lett.\  {\bf 92} (2004) 101802
  [arXiv:hep-ph/0308038].
    
\bibitem{Belyaev:2009ve}
  A.~S.~Belyaev, R.~S.~Chivukula, N.~D.~Christensen, E.~H.~Simmons, H.~J.~He, M.~Kurachi and M.~Tanabashi,
  arXiv:0907.2662 [hep-ph].
    
\bibitem{Accomando:2008dm}
  E.~Accomando, S.~De Curtis, D.~Dominici and L.~Fedeli,
  Nuovo Cim.\  {\bf 123B} (2008) 809
  [arXiv:0807.2951 [hep-ph]].

\bibitem{Casalbuoni:1985kq}
  R.~Casalbuoni, S.~De Curtis, D.~Dominici and R.~Gatto,
  Phys.\ Lett.\  B {\bf 155} (1985) 95.
    
\bibitem{Barbieri:2003pr}
  R.~Barbieri, A.~Pomarol and R.~Rattazzi,
  Phys.\ Lett.\  B {\bf 591} (2004) 141
  [arXiv:hep-ph/0310285].

\bibitem{Cacciapaglia:2004jz}
  G.~Cacciapaglia, C.~Csaki, C.~Grojean and J.~Terning,
  Phys.\ Rev.\  D {\bf 70} (2004) 075014
  [arXiv:hep-ph/0401160].

\bibitem{Cacciapaglia:2004rb}
  G.~Cacciapaglia, C.~Csaki, C.~Grojean and J.~Terning,
  Phys.\ Rev.\  D {\bf 71} (2005) 035015
  [arXiv:hep-ph/0409126].
  
\bibitem{Foadi:2004ps}
  R.~Foadi, S.~Gopalakrishna and C.~Schmidt,
  Phys.\ Lett.\  B {\bf 606}  (2005) 157
  [arXiv:hep-ph/0409266].
    
\bibitem{Dawson:2008as}
  S.~Dawson and C.~B.~Jackson,
  Phys.\ Rev.\  D {\bf 79} (2009) 013006
  [arXiv:0810.5068 [hep-ph]].

\bibitem{Cacciapaglia:2007fw}
  G.~Cacciapaglia, C.~Csaki, J.~Galloway, G.~Marandella, J.~Terning and A.~Weiler,
  JHEP {\bf 0804} (2008) 006
  [arXiv:0709.1714 [hep-ph]].

\bibitem{Cacciapaglia:2005pa}
  G.~Cacciapaglia, C.~Csaki, C.~Grojean, M.~Reece and J.~Terning,
  Phys.\ Rev.\  D {\bf 72} (2005) 095018
  [arXiv:hep-ph/0505001].
        
\bibitem{Agashe:2006at}
  K.~Agashe, R.~Contino, L.~Da Rold and A.~Pomarol,
  Phys.\ Lett.\  B {\bf 641} (2006) 62
  [arXiv:hep-ph/0605341].
  
\bibitem{Cacciapaglia:2006gp}
  G.~Cacciapaglia, C.~Csaki, G.~Marandella and J.~Terning,
  Phys.\ Rev.\  D {\bf 75} (2007) 015003
  [arXiv:hep-ph/0607146].

\bibitem{DeRoeck:2009id}
  A.~De Roeck {\it et al.},
  arXiv:0909.3240 [hep-ph].
      
\bibitem{Birkedal:2004au}
  A.~Birkedal, K.~Matchev and M.~Perelstein,
  Phys.\ Rev.\ Lett.\  {\bf 94} (2005) 191803
  [arXiv:hep-ph/0412278].
  
\bibitem{Englert:2008wp}
  C.~Englert, B.~Jager and D.~Zeppenfeld,
  JHEP {\bf 0903} (2009) 060
  [arXiv:0812.2564 [hep-ph]].

\bibitem{Martin:2009gi}
  A.~Martin and V.~Sanz,
  arXiv:0907.3931 [hep-ph].
  
\bibitem{Georgi:2007ek}
  H.~Georgi,
  Phys.\ Rev.\ Lett.\  {\bf 98} (2007) 221601
  [arXiv:hep-ph/0703260].

\bibitem{Strassler:2006im}
  M.~J.~Strassler and K.~M.~Zurek,
  Phys.\ Lett.\  B {\bf 651} (2007) 374
  [arXiv:hep-ph/0604261].
  
\bibitem{Kang:2008ea}
  J.~Kang and M.~A.~Luty,
  arXiv:0805.4642 [hep-ph].
  
\bibitem{HiggsSpace}
C.~Jackson, G.~Servant, G.~Shaughnessy, T.~M.~P.~Tait and M.~Taoso,
to appear. 
  
\bibitem{Grojean:2004xa}
  C.~Grojean, G.~Servant and J.~D.~Wells,
  Phys.\ Rev.\  D {\bf 71} (2005) 036001
  [arXiv:hep-ph/0407019].
  
\bibitem{Grojean:2006bp}
  C.~Grojean and G.~Servant,
  Phys.\ Rev.\  D {\bf 75} (2007) 043507
  [arXiv:hep-ph/0607107].
  
  
  


  


  

  
       
      
\end{thebibliography}
\end{document}